\documentclass[acmsmall]{acmart}
\AtBeginDocument{%
  }
\usepackage{algorithm}
\usepackage{algpseudocode}
\usepackage{longtable}
\usepackage{graphicx}
\usepackage{float}
\graphicspath{ {./graphs/} }
\usepackage{tabularx}
\usepackage{lscape} 
\setcopyright{acmlicensed}
\copyrightyear{2025}
\acmYear{2025}

\begin{document}
\title{A Survey on Task Scheduling in Carbon-Aware Container Orchestration}

\author{Jialin Yang}
\authornotemark[1]
\email{jialin.yang@ucalgary.ca}
\orcid{0009-0000-4875-1047}
\affiliation{%
  \institution{Department of Electrical and Software Engineering, University of Calgary}
  \streetaddress{2500 University Drive NW}
  \city{Calgary}
  \state{Alberta}
  \country{Canada}
  \postcode{T2N 1N4}
}

\author{Zainab Saad}
\email{zainab.saad1@ucalgary.ca}
\affiliation{%
  \institution{Department of Electrical and Software Engineering, University of Calgary}
  \streetaddress{2500 University Drive NW}
  \city{Calgary}
  \state{Alberta}
  \country{Canada}
  \postcode{T2N 1N4}
}

\author{Jiajun Wu}
\email{jiajun.wu1@ucalgary.ca}
\orcid{0000-0003-1000-7356}
\affiliation{%
  \institution{Department of Electrical and Software Engineering, University of Calgary}
  \streetaddress{2500 University Drive NW}
  \city{Calgary}
  \state{Alberta}
  \country{Canada}
  \postcode{T2N 1N4}
}

\author{Xiaoguang Niu}
\email{xgniu@whu.edu.cn}
\orcid{0000-0003-4252-3291}
\affiliation{%
  \institution{School of Computer Science, Wuhan University}
  \streetaddress{Luojiashan Road, Wuchang District}
  \city{Wuhan}
  \state{Hubei}
  \country{China}
  \postcode{430079}
}

\author{Henry Leung}
\affiliation{%
  \institution{Department of Electrical and Software Engineering, University of Calgary}
  \streetaddress{2500 University Drive NW}
  \city{Calgary}
  \state{Alberta}
  \country{Canada}}
\email{leungh@ucalgary.ca}
\orcid{0000-0002-5984-107X}

\author{Steve Drew}
\authornotemark[1]
\email{steve.drew@ucalgary.ca}
\orcid{0000-0003-4527-2635}
\affiliation{%
  \institution{Department of Electrical and Software Engineering, University of Calgary}
  \streetaddress{2500 University Drive NW}
  \city{Calgary}
  \state{Alberta}
  \country{Canada}
  \postcode{T2N 1N4}
}

\renewcommand{\shortauthors}{Jialin et al.}

\begin{abstract}

The soaring energy demands of large-scale software ecosystems and cloud data centers, accelerated by the intensive training and deployment of large language models, have driven energy consumption and carbon footprint to unprecedented levels.
In response, both industry and academia are increasing efforts to reduce the carbon emissions associated with cloud computing through more efficient task scheduling and infrastructure orchestration.
In this work, we present a systematic review of various Kubernetes scheduling strategies, categorizing them into hardware-centric and software-centric, annotating each with its sustainability objectives, and grouping them according to the algorithms they use. 
We propose a comprehensive taxonomy for cloud task scheduling studies, with a particular focus on the environmental sustainability aspect. 
We analyze emerging research trends and open challenges, and our findings provide critical insight into the design of sustainable scheduling solutions for next-generation cloud computing systems.
\end{abstract}

\ccsdesc[500]{General and reference~Surveys and overviews}
\ccsdesc[500]{General and reference~Reference works}

\keywords{ Container Orchestration, Kubernetes, Virtualization, Task Scheduling, Carbon Emissions}

\received{}
\received[revised]{}
\received[accepted]{}

\maketitle

\section{Introduction}\label{sec:intro}

With the globalization of digital services, cloud environments are now widely adopted across various industries~\cite{hardwareLowConsumption}. The three major cloud service providers, Amazon Web Services (AWS), Microsoft Azure, and Google Cloud Platform (GCP), deliver infrastructure and services on a global scale~\cite{sabastian2021international}. 
Driven in part by the rapid deployment of large language models (LLMs), the number and size of data centers have grown significantly, increasing electricity demand and associated carbon emissions~\cite{DataCenterHighConsumption}. 

In response to the environmental concerns, major cloud providers have begun to monitor, reduce, and publicly report their carbon footprints, aligning with broader sustainability goals ~\cite{patel2025sustainable}. For instance, AWS provides a comprehensive Customer Carbon Footprint Tool via its Sustainability Center\footnote{\url{https://sustainability.aboutamazon.com/products-services/aws-cloud}}. Microsoft offers the Emissions Impact Dashboard for Azure to estimate greenhouse gas emissions associated with cloud operations\footnote{\url{https://www.microsoft.com/en-us/sustainability}}. Similarly, GCP supplies a Carbon Footprint dashboard to reveal project-level emissions to customers\footnote{\url{https://cloud.google.com/carbon-footprint}}.

As containerized workloads become the dominant cloud-native paradigm, the efficiency of container orchestration directly impacts data center energy use and carbon emissions~\cite{piontek2024carbon}. As the leading container orchestration platform, Kubernetes offers a rich scheduling framework that can be extended to incorporate environmental metrics~\cite{rao2024energy}. Consequently, carbon-aware Kubernetes scheduling has emerged as a practical solution to improve the sustainability of cloud operations~\cite{dong2025towards}.

To understand its significance, it is necessary to start with virtualization and explore the technological foundations that support modern cloud infrastructure. Virtualization provides an abstraction layer between hardware and software by creating virtual instances of physical components such as CPUs, memory, networks, and storage~\cite{singh2018virtualization}. While virtualization improves efficiency, hypervisor-based virtual machines introduce additional overhead by running separate guest operating systems. Containerization offers a lightweight alternative by packaging application binaries and dependencies together while sharing the host OS kernel, enabling better resource utilization and faster deployment~\cite{bhardwaj2021virtualization}. Kubernetes provides built-in mechanisms to ensure high availability, scalability, and efficient workload scheduling, making it highly effective for managing large-scale, microservice-based applications~\cite{KubernetesContainers}.



To address these challenges, it is essential to further investigate existing cloud container scheduling algorithms within orchestration systems. Kubernetes scheduler, in particular, has shown strong potential in reducing carbon emissions by enabling efficient workload placement~\cite{carrion2022kubernetes}. 

To achieve seamless operation and management of containerized applications, a container orchestration engine is essential. It is responsible for scheduling and automatically scaling containers based on dynamic workload requirements. Designing an effective orchestration system requires the integration of multiple scheduling strategies, each optimized for specific objectives, such as minimizing response time, optimizing energy consumption, and maximizing resource utilization~\cite{Buyya_2018}. These strategies ensure the optimal allocation of physical resources among containers while simplifying the management and operation of large-scale containerized environments~\cite{rajuroy2025optimizing,raith2024opportunistic}.

\begin{table*}[htbp]
\scriptsize
\centering
\caption{Surveys and Reviews of Kubernetes or Container Scheduling. For each aspect, Both in \textit{Optimization} indicates the survey includes both hardware and software level optimization methods, \textit{Sustainability} considers environmental sustainability, \textit{Algorithmic labels} means if survey categorizes approaches based on algorithmic strategies, or \textit{Taxonomy Depth} indicates survey presents a well-structured, reusable classification framework.}
\begin{tabular}{p{1.6cm} p{0.5cm} p{3.8cm} p{1.9cm} p{1.3cm} p{1.3cm} p{1cm}}
\toprule
\textbf{Survey} & \textbf{Year} & \textbf{Focus} & \textbf{Optimization} & \textbf{Sustainability} & \textbf{Algorithmic Labels} & \textbf{Taxonomy Depth} \\
\midrule
Burns \textit{et al.}~\cite{burns2016borg} & 2016 & Lessons from a decade of Google’s container-management systems (Borg, Omega, Kubernetes), focusing on software design and scheduling evolution. & Software & No & No & No \\
Pahl \textit{et al.}~\cite{pahlSurvey}& 2017 & Systematic mapping of cloud container orchestration research, highlighting architecture, management, and technology concerns. & Software & No & Yes & Yes \\
Maria \textit{et al.} ~\cite{mariaSurvey} & 2018 & Taxonomy of software-based container orchestration mechanisms for scalability, fault tolerance, and resource efficiency. & Software & No & Yes & Yes \\
Hong \textit{et al.}~\cite{hongSurvey} & 2019 & Survey of resource management techniques in fog/edge systems including task placement, scheduling, and migration & Software & No & No & No \\
Casalicchio~\cite{Casalicchio2019Survey} & 2019 & Overview of Kubernetes, Docker Swarm, and container orchestration features & Software & No & No & No \\
Ahmad \textit{et al.}~\cite{ahmad2022containerSurvey} & 2022 & Classification of container scheduling algorithms including mathematics and heuristics etc. & Software & No & Yes & Yes \\
Carrión~\cite{carrion2022kubernetes} & 2022 & Taxonomy of Kubernetes scheduling and challenges in performance and energy. & Software & Yes & No & Yes \\
Rejiba \textit{et al.}~\cite{zeineb2022Survey} & 2022 & Survey of custom Kubernetes schedulers for diverse workloads, with taxonomy based on scheduling objectives, workloads, and environments. & Software & No & Yes & Yes \\
Senjab \textit{et al.}~\cite{Jawaddi2023Survey} & 2023 & Survey of Kubernetes scheduling algorithms categorized into generic, multi-objective, AI-based, and auto scaling approaches. & Software & No & Yes & No \\
\textbf{Ours} & \textbf{2025} & \textbf{Survey of sustainability-focused Kubernetes scheduling} & \textbf{Both} & \textbf{Yes} & \textbf{Yes} & \textbf{Yes} \\
\bottomrule
\end{tabular}
\label{tab:k8s-surveys}
\end{table*}

We present an overview of existing surveys in Kubernetes and container scheduling literature in Table \ref{tab:k8s-surveys}, categorizing them by optimization methods (software/hardware), consideration of sustainability goals, algorithmic labeling approaches, and taxonomy reusable classification framework.

Earlier surveys predominantly focus on software-based optimization methods, primarily due to their emphasis on performance metrics while largely ignoring hardware-level energy costs or industry insights. Moreover, few existing works have explicitly prioritized sustainability aspects such as environmental impacts, energy efficiency, or carbon-aware scheduling strategies.

Burns \textit{et al.} ~\cite{burns2016borg} provided foundational lessons from Google's container management systems, focusing mainly on software design without explicitly addressing sustainability. Pahl \textit{et al.} ~\cite{pahlSurvey} systematically mapped cloud container orchestration but did not consider environmental factors. Maria \textit{et al.} ~\cite{mariaSurvey} and Ahmad \textit{et al.} ~\cite{ahmad2022containerSurvey} proposed taxonomies focusing on scalability, resource efficiency, and scheduling algorithms but without addressing sustainability or integrating hardware considerations.

Recent surveys like Carrión~\cite{carrion2022kubernetes} and Senjab \textit{et al.}~\cite{Jawaddi2023Survey} started incorporating sustainability considerations. However, these are still primarily software-focused and lack significant industrial integration.

Compared with previous surveys, our survey significantly makes the following key contributions:

\begin{itemize}
    \item We systematically review cloud task scheduling strategies by incorporating both \textbf{hardware- and software-level optimization methods}, explicitly addressing \textbf{environmental sustainability} and integrating insights from both academic research and \textbf{industry practices}.
    
    \item We propose the \textbf{first systematic classification} of Kubernetes scheduling algorithms based on two dimensions: underlying optimization methods (hardware-driven or software-driven) and sustainability goals (energy efficiency or carbon emissions awareness).
    
    \item We introduce a \textbf{dual-layer taxonomy} that links technical scheduling strategies to their environmental impacts, bridging critical gaps in existing literature and offering \textbf{comprehensive guidance} for sustainability-focused scheduling in cloud environments.
\end{itemize}

\section{Survey Methodology}\label{sec:research}
To thoroughly investigate how academia and industry can utilize Kubernetes scheduling to reduce carbon intensity, we explored systematic evaluation methods such as MOOSE (Meta-Analysis of Observational Studies in Epidemiology)~\cite{Moose}, Cochrane Handbook for Systematic Reviews of Interventions~\cite{cochrane}, ROBIS (Risk Of Bias In Systematic Reviews)~\cite{robis}, and GRADE (Grading of Recommendations Assessment, Development, and Evaluation)~\cite{GRADE}. We selected PRISMA (Preferred Reporting Items for Systematic Reviews and Meta-Analyses)~\cite{prisma} because it is widely recognized and comprehensive. MOOSE is less recognized, the Cochrane Handbook is more suited for medical research, ROBIS lacks a complete framework, and GRADE’s assessments can be subjective. PRISMA provides a robust framework for our systematic review needs.

\begin{figure}[ht]
    \centering
    \includegraphics[scale=0.40]{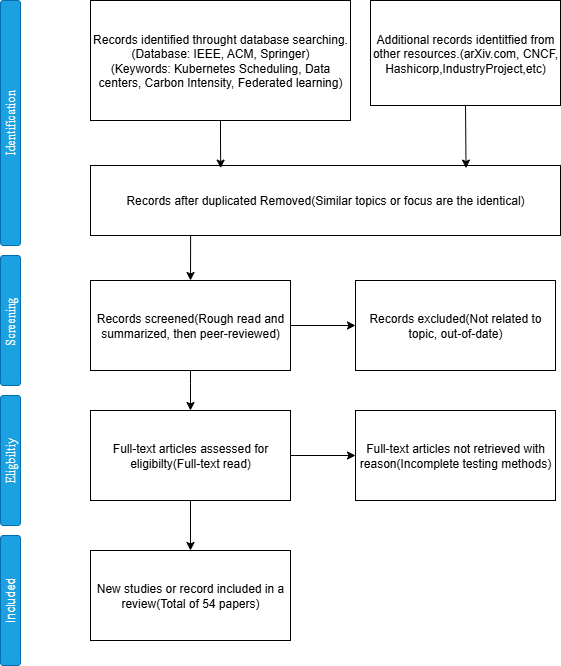}
    \Description{The PRISMA Flow Diagram to Collect papers}
    \caption{The PRISMA Flow Diagram to Collect papers}
    \label{fig:PRISMADIAGRAM}
\end{figure}

\subsection{Search Objectives}
In our review paper, we aim to explore the development of cloud computing and its implications, along with potential solutions to emerging challenges. Our research objectives are as follows:
\begin{itemize}
    \item[-]Outline the evolution of networks from virtualization to containerization and eventually to cloud computing.
    \item[-]Optimize orchestration engines, such as Kubernetes, to reduce carbon emissions.
    \item[-]Investigate the application of federated learning in scheduling optimization to further reduce carbon emissions.
\end{itemize}

\subsection{Search Strategy}
We conducted a comprehensive literature search using academic databases such as the ACM Digital Library, the Institute of Electrical and Electronics Engineers (IEEE) Xplore, and SpringerLink. Additionally, we utilized Google Scholar and Connected Papers as external tools to identify relevant works through citation mapping. Our primary search terms included “virtualization,” “Kubernetes scheduling,” “carbon intensity,” and “federated learning.” Connected Papers was particularly useful in identifying relevant research clusters based on citation distance.

To complement academic sources, we also incorporated gray literature and reports from major industry stakeholders, including Google, Microsoft, and the Green Software Foundation (GSF). This combination of academic and industry sources ensures that our review covers both foundational research and the latest practical innovations related to sustainable cloud orchestration.

\subsection{Eligibility Criteria}
To maintain the quality and relevance of this review, we applied the following inclusion and exclusion criteria during the screening process illustrated in \autoref{fig:PRISMADIAGRAM}.
\subsubsection{Inclusion Criteria}
\begin{itemize}
    \item The study is published in a high-impact, peer-reviewed journal or top-tier conference proceedings.
    \item The content aligns with core topics such as virtualization, Kubernetes scheduling, carbon efficiency, and federated learning.
    \item The methodology presented is rigorous and well-documented.
\end{itemize}
\subsubsection{Exclusion Criteria}
\begin{itemize}
    \item The study's focus falls outside the scope of sustainable Kubernetes scheduling or cloud optimization.
    \item The source credibility is questionable or unverifiable.
    \item The publication is outdated, unless it represents a seminal or highly cited foundational work.
\end{itemize}
This systematic filtering approach ensures that only the most relevant and high-quality studies are included in our analysis.

\section{Preliminary}\label{sec:background}
The rapid adoption of Kubernetes for container orchestration has increased the urgency of addressing environmental sustainability issues, particularly the energy consumption and resource allocation challenges of emerging data centers. In addition, concerns about data privacy preservation between distributed devices have highlighted the importance of federated learning as a potential solution.

\subsection{Energy Consumption Challenge}
Efficient scheduling in Kubernetes is directly related to broader energy consumption challenges. Data centers, central to cloud and containerized workloads, significantly contribute to global energy use and emissions. From 2010–2018, global data center electricity grew by 6\%, reaching ~1\% of global consumption, with server utilization and storage efficiency also increasing~\cite{KubernetesContainers, DCEnergyUse}. However, data centers now exhibit higher energy intensity per GDP unit than other industries.

In the U.S., server shipments grew from 3M (2016) to over 6.5M (2022), driven by GPU-heavy AI workloads, pushing electricity consumption from 60 TWh to 176 TWh at 2023 which is 4.4\% of national demand, according to the projections, it may up to 580 TWh (12\%) by 2028~\cite{USDataCenterCount} .Ireland expects data center energy use to hit 32\% of its national electricity by 2026~\cite{IEAreport}. Globally, data center use may exceed 1,000 TWh by 2026, comparable to Japan’s total use. Efficient, carbon-aware scheduling is crucial for mitigating this growth~\cite{resourceallocationreview}.

\subsection{Resource Allocation}
In order to relieve energy consumption challenges, it is crucial to optimize resource allocation. Traditional cost-based scheduling aims to minimize task time and operational cost~\cite{banga2020cost}. Modern schedulers use heuristic and meta-heuristic algorithms~\cite{mor2021heuristic, jain2017cloud} to address multi-resource allocation challenges in virtualized environments~\cite{rodriguez2019container}.

\textbf{Power Management:} Techniques like sleep states, DVFS, and geo-distributed load balancing help reduce energy use. For example, Zheng and Cai propose a model optimizing power cost across regions with variable electricity prices~\cite{ZHENG2011275}.

\textbf{Resource Management:} Adaptive workload scheduling strategies like CPU scaling, consolidation, and asynchronous offloading reduce active nodes and enhance efficiency~\cite{quan2012t}.

\textbf{Thermal Management:} Efficient thermal strategies reduce cooling overhead. Metrics by Lajevardi enable thermal-aware scheduling~\cite{LAJEVARDI2015511}, while Beitelmal shows that higher CPU thermal thresholds can enhance energy efficiency~\cite{BEITELMAL2014562}.

These dimensions jointly optimize computation, cost, and energy in sustainable data center operation.

\subsection{Federated Learning}
Given that resource allocation often involves distributed and private data, federated learning has emerged as a promising method that balances privacy and efficiency. Federated learning (FL) enables decentralized model training, preserving privacy and reducing communication overhead. It is especially useful in energy domains where renewable sources like wind and solar vary across time and space. Accurate prediction models require diverse data, yet the industry hesitates to share it~\cite{flgreenreview}.

Kusiak highlights that high-resolution turbine data is often inaccessible due to manufacturer restrictions~\cite{kusiak2016renewables}. FL overcomes this barrier by enabling collaborative learning without centralizing data. Studies show FL achieves comparable accuracy to centralized models across various green energy applications~\cite{li2023wind, ahmadi2022deep, wang2022privacy, wang2023efficient, moayyed2022cyber}.

\subsection{Carbon Awareness}
To effectively utilize federated learning for sustainable computing, it is crucial to raise carbon awareness among stakeholders. Carbon awareness emphasizes understanding and mitigating CO\textsubscript{2} emissions. Raising awareness supports behavioral change and adoption of sustainable technologies~\cite{Gevrek2015Public}.

Carbon taxes impose costs on emitters to incentivize greener practices~\cite{Timilsina2022Carbon, Baranzini2000A}. Revenue may fund renewable energy or offset public costs. Data centers, which could account for 8\% of global emissions by 2030~\cite{Cao2021Toward}, are prime targets. Carbon taxation can drive adoption of clean energy and energy-efficient infrastructure~\cite{Bosse2020Quantitative}.

\subsection{Kubernetes Default Scheduler}
Addressing the above carbon emission awareness issues requires a reconsideration of Kubernetes, the mainstream container orchestration framework, especially its scheduling strategy. The \textbf{Kube-scheduler} is the default scheduler component of Kubernetes. It is responsible for assigning unscheduled Pods to appropriate Nodes based on a combination of policies, constraints, and resource availability. Its scheduling algorithm follows a two-phase design: \textit{filtering} and \textit{scoring}~\cite{sigelman2019kubernetes,k8s-scheduler-concepts}. 

\subsubsection*{Filtering Phase}

In the filtering phase, the scheduler eliminates Nodes that do not satisfy basic scheduling requirements~\cite{k8s-scheduling-framework,k8s-scheduler-concepts}. Some default filters (also called \textit{predicates}) include:

\begin{itemize}
    \item \texttt{PodFitsResources}: Verifies if the Node has sufficient CPU and memory.
    \item \texttt{NodeAffinity}: Checks if the Pod's node affinity rules match the Node.
    \item \texttt{TaintsAndTolerations}: Ensures that the Pod can tolerate the Node’s taints.
    \item \texttt{VolumeBinding}: Ensures that required volumes can be mounted on the Node.
\end{itemize}

Only Nodes that pass all filtering criteria are considered for the next phase~\cite{verma2015large}.

\subsubsection*{Scoring Phase}

In the scoring phase, each filtered Node is assigned a score, and the Node with the highest score is selected~\cite{k8s-scheduler-concepts}. Some common built-in scoring functions (also called \textit{priorities}) include:

\begin{itemize}
    \item \texttt{Least-allocated}: Prefers Nodes with the most available resources, promoting balanced resource usage.
    \item \texttt{Most-allocated}: Opposite of \texttt{Least-allocated}; can help with resource bin-packing.
    \item \texttt{BalancedResourceAllocation}: Scores Nodes based on balanced CPU and memory allocation.
    \item \texttt{TopologySpreadConstraint}: Spreads Pods evenly across failure domains (e.g., zones or racks).
\end{itemize}

These scores are often weighted and normalized before a final decision is made. If multiple Nodes share the highest score, one random node is selected.

\subsubsection*{Disadvantages and Limitations}

While the \texttt{kube-scheduler} provides a general-purpose, efficient scheduling mechanism, it has several limitations:

\begin{itemize}
    \item \textbf{Resource-Centric Only:} Default scoring plugins focus primarily on CPU and memory. Other concerns like energy consumption, carbon intensity, or workload priorities are not considered unless customized~\cite{Resource-Centric}.
    
    \item \textbf{No Learning or Adaptation:} The default scheduler does not learn from previous scheduling outcomes or adapt to dynamic patterns in workloads~\cite{NoLearning}.
    
    \item \textbf{Limited Global View:} The scheduler makes decisions Pod-by-Pod and does not globally optimize across jobs or queues~\cite{limitedGoalView}.
    
    \item \textbf{Reactive, Not Predictive:} The scheduler reacts to Pod creation events, rather than predicting future load or preemptively optimizing node utilization~\cite{reactive}.
    
    \item \textbf{Random Tie-Breaking:} When multiple Nodes share the highest score, one is chosen randomly, which may lead to suboptimal or inconsistent scheduling behavior~\cite{randomTieBreaking}.
    
    \item \textbf{Inadequate for Specialized Policies:} Workloads requiring custom objectives (e.g., energy efficiency, data locality, SLA guarantees) must implement plugins or alternative schedulers~\cite{Inadequate}.
\end{itemize}

These limitations have led to a wide array of research and engineering efforts aimed at developing smarter, more context-aware schedulers, many of which integrate with or replace \texttt{kube-scheduler} through scheduling frameworks.

\section{Industry Efforts Toward Carbon-Aware Computing}
As the carbon footprint of data centers and containerized applications continues to grow, stakeholders across the industry have actively taken steps to address these sustainability challenges. Driven by the rapid growth in cloud computing and Kubernetes adoption, organizations in academia and industry are increasing their research and development efforts into carbon-aware computing solutions and sustainable scheduling practices.

The AI Alliance brings together top universities such as Imperial College London, Keio University, and Yale with major technology companies including ORACLE, IBM, and META to explore AI’s potential in reducing carbon emissions~\cite{AIAlliance}.

In 2021, the Cloud Native Computing Foundation launched \textbf{Crane}, an open-source Cloud Native FinOps project that enables Kubernetes users to monitor carbon footprints, analyze resource usage, and optimize workloads to reduce emissions.

The \textbf{Green Software Foundation}, a non-profit organization, developed the Carbon Aware SDK to measure software carbon impact and adjust runtime behavior to improve environmental sustainability~\cite{cncfCrane}.

\textbf{HashiCorp’s Nomad}, a multi-region orchestration engine, supports carbon-aware scheduling using location-based algorithms. It applies binpack scheduling to shift workloads to greener regions and adjusts task anti-affinity based on local carbon intensity to prevent node overload. This approach can be further improved by integrating task-specific characteristics~\cite{hashicorpNomad}.

According to the \textbf{BP Energy Outlook 2030}~\cite{BP2018Outlook}, renewable energy sources are projected to account for the largest share of global energy demand growth over the next two decades. This trend is progressing more rapidly than previous transitions such as nuclear and natural gas, signaling a shift in the energy landscape for traditional oil companies.

However, this transition remains fragmented. European firms such as Shell, Total, and Equinor have made substantial investments in renewable technologies like wind, solar, and electric vehicle infrastructure~\cite{sheppard2018oil, mackenzie2018shell}. In contrast, U.S.-based companies such as Exxon, Mobil and Chevron have focused primarily on improving fossil fuel efficiency with limited diversification into renewable~\cite{ward2018oil, Mobil}. Geothermal energy remains significantly under invested across all major players.

These industry-led efforts highlight the urgency of addressing energy consumption and carbon emissions, especially given the critical role of Kubernetes scheduling in cloud-native infrastructure. However, achieving truly sustainable scheduling practices requires addressing deeper, fundamental challenges, including energy efficiency, effective resource allocation, protecting data privacy through federated learning, and enhancing carbon awareness across technical platforms~\cite{USPolar}. Despite growing awareness and momentum, the lack of coordinated action and the diversity of energy strategies across regions and organizations continue to hinder a unified and effective transition to carbon-aware computing~\cite{BritishPolar}.

\section{Scheduling Algorithm} \label{sec:schedulingAlgo}
Research on container scheduling in cloud-native environments generally falls into two categories: \textbf{hardware-level optimization}, which leverages telemetry and external signals (e.g., energy sources, hardware status, data center traits), and \textbf{software-level algorithmic approaches}, which refine resource allocation and scheduling logic. These methods target either \textbf{energy efficiency} or \textbf{carbon awareness} as their primary sustainability goal.

To reflect the breadth of current techniques, we further classify scheduling strategies into the following groups: \textbf{Energy Consumption Priority Schedulers}, \textbf{Data Center Cost Management}, \textbf{Multi-Criteria Optimization}, \textbf{Temporal and Spatial Carbon Shifting}, \textbf{Carbon-Aware Workload Shifting}, \textbf{ML/DL-Based Scheduling}, \textbf{Heuristic and Metaheuristic Methods}, \textbf{Energy Consumption Metrics Monitoring}, \textbf{Fair Resource Allocation}, \textbf{AI-Driven and DRL Schedulers}, \textbf{Serverless, Edge, and Federated Solutions}, and \textbf{Carbon Emission Index Priority Scheduling}. Papers originating from the \textbf{industry sector} are marked with ``†'' to highlight practical contributions.

\textbf{Energy Efficiency} aims to reduce total power consumption by improving infrastructure utilization and minimizing idle resource usage without compromising system performance or quality of service. This goal is often pursued through fine-grained scheduling, load consolidation, or dynamic scaling techniques that minimize energy waste.

\textbf{Carbon Awareness}, on the other hand, prioritizes reducing greenhouse gas emissions by accounting for the carbon intensity of electricity used. It often involves time-shifting or geo-shifting workloads to align with periods or locations where cleaner (e.g., renewable) energy is available. While this may introduce trade-offs in terms of latency, cost, or resource utilization, it enables a lower-carbon operational footprint aligned with sustainability goals.

\begin{figure}[htbp]
\centering
\includegraphics[width=0.95\textwidth]{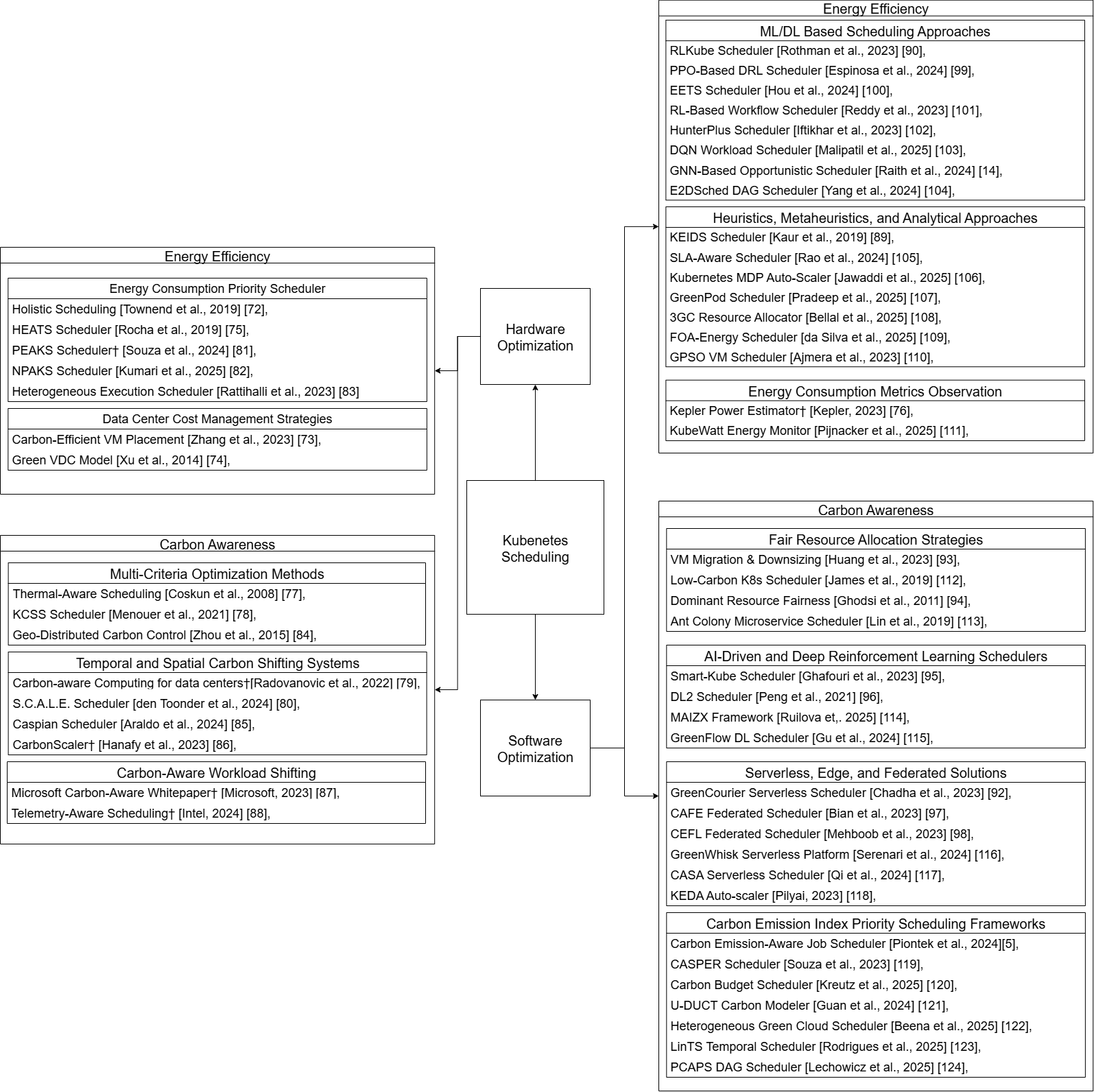}
\caption{Taxonomy of Kubernetes Scheduling Strategies Categorized by Optimization Types and Sustainability Objectives. Note: † Industry paper}
\Description{A hierarchical taxonomy diagram of Kubernetes scheduling strategies. }
\label{fig:taxonomy}
\end{figure}

\autoref{fig:taxonomy} presents an initial visual taxonomy of existing Kubernetes scheduling approaches, organized by whether they target hardware or software optimization, and further divided based on their sustainability goal of energy efficiency or carbon awareness. More detailed descriptions of these approaches are provided in the subsequent summary tables.

\subsection{Scheduling Overview}

\textbf{Hardware Optimization}
This category focuses on scheduling strategies that exploit telemetry and hardware-level signals to achieve sustainability goals. The proposals can be broadly categorized into energy efficiency and carbon awareness strategies.

For energy efficiency, notable proposals include a holistic scheduling algorithm that replaces Kubernetes' default scheduler by incorporating both physical and virtual infrastructure states and business logic ~\cite{townend2019improving}; a virtual data center management framework designed to reduce emissions and improve cloud infrastructure profitability ~\cite{xu2014virtual}; and methods for Virtual Machine placement and migration that reduce unnecessary energy consumption ~\cite{zhang2023carbon, rocha2019heats}. Frameworks such as Kepler provide real-time power usage estimates at the container level, further enhancing energy visibility and decision-making accuracy ~\cite{kepler}.

On the carbon awareness front, several strategies emphasize adjusting workloads based on carbon intensity data. Examples include temperature-aware task scheduling for chip-level SoCs(System on Chip) ~\cite{coskun2008static}, the Kubernetes Container Scheduling Strategy (KCSS) which employs multi-criteria decision-making to reduce emissions ~\cite{menouer2021kcss}, and Google’s Carbon-Intelligent Compute system that reschedules jobs based on predicted carbon intensity ~\cite{radovanovic2022carbon}. Carbon Scaler dynamically adjusts server allocations using real-time carbon data and Kubernetes auto-scaling to achieve up to 51\% emission reduction ~\cite{hanafy2023carbonscaler}.

\begin{table}[htbp]
\scriptsize
\caption{Hardware Optimization Techniques for Energy-Efficient Scheduling}
\noindent\textbf{\large Category: Energy Efficiency} \par\vspace{1ex}
\begin{tabularx}{\textwidth}{|p{4.2cm}|X|}
\hline
\textbf{Title} & \textbf{Key Proposal} \\ \hline

\multicolumn{2}{|c|}{\textbf{Energy Consumption Priority Scheduler}} \\ \hline
Improving Data Center Efficiency via Holistic Scheduling ~\cite{townend2019improving} & A \textbf{holistic scheduling algorithm} is used to replace Kubernetes' default scheduler. This new algorithm considers the impact of virtual and physical infrastructure and business processes. \\ \hline
HEATS: Heterogeneity-and Energy-Aware Task-Based Scheduling ~\cite{rocha2019heats} & Study the hardware architectures associated with schedulers and migrate them to different cluster nodes to meet the deployment carbon and performance tradeoffs. \\ \hline
PEAKS: Power-Efficiency-Aware Kubernetes Scheduler†~\cite{souza2024peaks} & \textbf{PEAKS} is a Kubernetes scheduler plugin that reduces power consumption by prioritizing pod placement on nodes with higher energy efficiency, as measured by real-time metrics from Kepler~\cite{kepler}. \\ \hline
Real-Time Node’s Power-Aware Kubernetes Scheduler(NPAKS) in a Cloud Environment ~\cite{kumari2025npaks} & \textbf{NPAKS} is an energy-aware Kubernetes scheduler that uses real-time node-level energy consumption data to reduce cluster energy consumption while maintaining performance. \\ \hline
Fine-Grained Heterogeneous Execution Framework with Energy Aware Scheduling ~\cite{rattihalli2023fine} & The authors propose a Kubernetes-based energy-aware scheduler that achieves significant reductions in energy consumption and execution time by optimizing \textbf{task co-location} and \textbf{resource utilization} on heterogeneous hardware. \\ \hline

\multicolumn{2}{|c|}{\textbf{Data Center Cost Management Strategies}} \\ \hline
A Virtual Data Center Deployment Model Based on the Green Cloud Computing ~\cite{xu2014virtual} & This paper proposes a virtual data center management framework with a purpose of reducing carbon emissions to increase the profits of cloud computing infrastructure suppliers and optimize infrastructure providers' external environment. \\ \hline
Carbon-Efficient Virtual Machine Placement in Cloud Data centers over Optical Networks ~\cite{zhang2023carbon} & Examine the relationship between data center configurations( such as latency, bandwidth, and capacity) and emissions to minimize carbon emissions without compromising service quality. \\ \hline
\end{tabularx}
\vspace{-1ex}
\begin{flushleft}
\scriptsize † Industry paper
\end{flushleft}
\end{table}

\begin{table}[htbp]
\scriptsize
\caption{Hardware Optimization Techniques for Carbon-Aware Scheduling}
\noindent\textbf{\large Category: Carbon Awareness} \par\vspace{1ex}
\begin{tabularx}{\textwidth}{|p{4.2cm}|X|}
\hline
\textbf{Title} & \textbf{Key Proposal} \\ \hline

\multicolumn{2}{|c|}{\textbf{Multi‑Criteria Optimization Methods}} \\ \hline
Static and Dynamic Temperature-Aware Scheduling for Multiprocessor SoCs ~\cite{coskun2008static} & The task scheduling problem is solved using \textbf{Integer Linear Programming(ILP)} under the premise of minimizing energy, balancing energy by and reducing hot spots, and designing dynamic scheduling strategies at the software level, which drastically reduces high-intensity thermal cycling. \\ \hline
KCSS: Kubernetes Container Scheduling Strategy ~\cite{menouer2021kcss} & \textbf{KCSS} uses the \textbf{TOPSIS algorithm} to find the optimal algorithm to reduce carbon emissions when introducing CPU, memory, disk utilization rate, power consumption, number of containers, and image transmit time. \\ \hline
Carbon-Aware Online Control of Geo-Distributed Cloud Services ~\cite{zhou2015carbon} & This study presents a framework that exploits \textbf{spatial} and \textbf{temporal variability} in electricity carbon footprints to reduce emissions in geo-distributed cloud services. The approach uses \textbf{Lyapunov optimization} for online decision-making regarding load balancing, capacity right-sizing, and server speed scaling to minimize electricity costs and emissions simultaneously. \\ \hline

\multicolumn{2}{|c|}{\textbf{Temporal and Spatial Carbon Shifting Systems}} \\ \hline

Carbon-Aware Computing for Data centers† ~\cite{radovanovic2022carbon} & \textbf{Google}'s Carbon-Intelligent Compute Management system minimizes the carbon footprint by scheduling flexible workloads during periods of lower carbon intensity. This system uses day-ahead carbon intensity forecasts and adjusts the compute capacity accordingly. \\ \hline
CarbonScaler: Leveraging Cloud Workload Elasticity for Optimizing Carbon-Efficiency ~\cite{hanafy2023carbonscaler} & \textbf{CarbonScaler} dynamically adjusts server allocation based on real-time carbon intensity data, significantly reducing emissions for cloud workloads. The system uses Kubernetes' autoscaling capabilities to manage batch jobs, achieving up to 51\% carbon savings over noncarbon-aware approaches. \\ \hline
Caspian – Carbon-Aware Multi-Cluster Scheduling for Cloud-Native Workloads† ~\cite{ibm2024caspian} & \textbf{Caspian}, developed by IBM,  a Kubernetes scheduler that reduces carbon emissions by dynamically shifting workloads across regions and times with lower grid carbon intensity. \\ \hline
S.C.A.L.E.: Scheduler for Carbon-Aware Load Execution at ING† ~\cite{denToonder2024scale} & \textbf{S.C.A.L.E.} is a carbon-aware batch scheduler that reschedules jobs in Kubernetes to greener time windows based on carbon intensity forecasts, reducing emissions while meeting workload deadlines. \\ \hline

\multicolumn{2}{|c|}{\textbf{Carbon‑Aware Workload Shifting}} \\ \hline
Microsoft Carbon-Aware Computing White paper† ~\cite{Microsoft2023} & Shift loads according to time and place to avoid heavy use of dirty energy sources such as fossil fuels. \\ \hline

Intel Telemetry Aware Scheduling†~\cite{intel2024tas} & \textbf{Telemetry Aware Scheduling (TAS)} by Intel enables Kubernetes to make smarter pod placement decisions by integrating fine-grained hardware telemetry. \\ \hline

\end{tabularx}
\vspace{-1ex}
\begin{flushleft}
\scriptsize † Industry paper
\end{flushleft}
\end{table}

\textbf{Software Optimization}

This category targets algorithmic-level enhancements to container orchestration systems to meet energy and carbon objectives. These approaches are likewise divided into energy efficiency and carbon awareness.

For energy efficiency, methods such as KEIDS apply integer linear programming to optimize task scheduling in edge-cloud IoT environments while minimizing power usage ~\cite{kaur2019keids}. RLKube uses reinforcement learning to train a Double DQN with Prioritized Experience Replay to learn optimized Kubernetes scheduling policies ~\cite{rothman2023rl}. Other approaches incorporate AI-based decision models for container placement to reduce power consumption ~\cite{jorge2021artificial}.

Carbon awareness strategies include dynamic workload shifting based on regional carbon intensity (e.g., GreenCourier for serverless platforms ~\cite{chadha2023greencourier}), algorithms that detect idle or under-utilized VMs and downsize them to reduce emissions ~\cite{huang2023reducing}, and fairness-based resource allocation to prevent carbon-heavy overload on specific nodes ~\cite{ghodsi2011dominant}. Deep learning-based schedulers like Smart-Kube and DL2 further demonstrate how intelligent resource control can strike a balance between performance and sustainability ~\cite{ghafouri2023smart, peng2021dl2}. Federated learning schedulers such as CAFE and CEFL promote privacy-preserving and carbon-efficient training in distributed environments by keeping data local and optimizing compute timing ~\cite{cafe, cefl}.

\begin{table}[htbp]
\scriptsize
\caption{Summary of Energy-Efficient Scheduling Techniques in Kubernetes and Cloud Environments}
\noindent\textbf{\large Category: Energy Efficiency} \par\vspace{1ex}
\begin{tabularx}{\textwidth}{|p{4.5cm}|X|}
\hline
\textbf{Title} & \textbf{Key Proposal} \\ \hline

\multicolumn{2}{|c|}{\textbf{Machine Learning and Deep Learning Based Scheduling Approaches}} \\ \hline
An RL-Based Model for Optimized Kubernetes Scheduling ~\cite{rothman2023rl} & \textbf{RLKube}, a custom Kubernetes scheduler using Reinforcement Learning, improves task scheduling efficiency, enhancing energy efficiency and resource utilization by utilizing Double Deep Q-Network with Prioritized Experience Replay. \\ \hline
Optimizing Energy Consumption of Kubernetes Clusters with Deep Reinforcement Learning ~\cite{espinosa2024optimizing} & This paper introduces a \textbf{Proximal Policy Optimization(PPO)} based \textbf{Deep Reinforcement Learning(DRL)} scheduler integrated into Kubernetes that reduces energy consumption by up to 24\% through intelligent pod placement in edge-cloud environments. \\ \hline
EETS: An energy-efficient task scheduler in cloud computing based on improved DQN algorithm ~\cite{hou2024eets} & \textbf{EETS} is a deep reinforcement learning based task scheduling algorithm that combines Double Dueling DQN and prioritized experience replay to minimize energy consumption and task response time in cloud environments. \\ \hline
An energy efficient RL based workflow scheduling in cloud computing ~\cite{reddy2023energy} & This paper proposes a secure, energy-efficient, \textbf{RL-based} workflow scheduling framework for cloud computing that outperforms existing approaches in terms of reduced energy consumption, security, cost and time horizon. \\ \hline
HunterPlus: AI based energy-efficient task scheduling for cloud--fog computing environments ~\cite{iftikhar2023hunterplus} & \textbf{HunterPlus} is a deep learning based task scheduling algorithm for cloud and fog environments that optimizes workload placement to reduce energy consumption by 17\% and improves the job completion rates by 10.4\%. \\ \hline
Energy-Efficient Cloud Computing Through Reinforcement Learning-Based Workload Scheduling ~\cite{malipatil2025energy} & The paper uses \textbf{Deep Q-Networks (DQN)} to learn an energy-efficient workload scheduling policy, using feature engineered inputs such as execution time, CPU  and memory consumption and makes informed scheduling decisions that optimize both energy efficiency and performance. \\ \hline
Opportunistic Energy-Aware Scheduling for Container Orchestration Platforms Using Graph Neural Networks ~\cite{raith2024opportunistic} & This paper proposes a \textbf{Graph Neural Network (GNN)} model to predict power consumption with a root mean square error of 7.5\%, along with a set of opportunistic scheduling algorithms to schedule applications based on the GNN estimated energy impact of incoming containers. The scheduler achieves a 6.2\% reduction in energy consumption. \\ \hline
Energy-efficient DAG scheduling with DVFS for cloud data centers ~\cite{yang2024energy} & This paper presents \textbf{E2DSched}, an online reinforcement learning based energy-efficient task scheduling algorithm for randomly arriving directed acyclic (DAG) graph jobs in cloud data centers. \\ \hline

\multicolumn{2}{|c|}{\textbf{Heuristics, Metaheuristics, and Analytical Approaches}} \\ \hline
KEIDS: Kubernetes-Based Energy and Interference Driven Scheduler for Industrial IoT in Edge-Cloud Ecosystem ~\cite{kaur2019keids} & The \textbf{KEIDS} scheduler minimizes energy consumption and carbon emissions in edge-cloud nodes for IoT setups by optimizing task scheduling using integer linear programming for multiobjective optimization. \\ \hline

Energy-aware Scheduling Algorithm for Microservices in Kubernetes Clouds ~\cite{rao2024sla} & This paper proposes an energy-efficient Kubernetes scheduling algorithm based on \textbf{service level agreements (SLAs)} that reduces cluster energy consumption by optimizing pod placement using communication-aware prioritization and an improved Sparrow Search meta-heuristic algorithm. \\ \hline

Analyzing Energy‑Efficient and Kubernetes‑Based Autoscaling of Microservices Using Probabilistic Model Checking ~\cite{Jawaddi2025} & The paper introduces a formal model-checking framework using \textbf{Markov Decision Process(MDP)} to evaluate and optimize energy efficiency in Kubernetes auto scaling policies for microservice workloads. \\ \hline
GreenPod: Energy-Optimized Scheduling for AIoT Workloads Using TOPSIS ~\cite{pradeep2025energy} & \textbf{Greenpod} is a multi-criteria Kubernetes scheduler that uses TOPSIS to optimize energy efficiency in (Artificial Intelligence of Things)AIoT workloads, achieving up to 39.1\% energy savings over the default scheduler. \\ \hline
3GC: A deadline-aware and energy-efficient resource allocation scheme for serverless edge computing ~\cite{bellal20253gc} & \textbf{3GC} is a deadline-aware resource allocation strategy for serverless edge computing that minimizes energy and execution costs using real-time allocation and \textbf{Dynamic Voltage and Frequency Scaling(DVFS)-based} tuning while ensuring latency constraints. \\ \hline
FOA-Energy: A Multi-objective Energy-Aware Scheduling Policy for Serverless-based Edge-Cloud Continuum ~\cite{da2025foa} & This paper presents \textbf{FOA-Energy}, a two-level multi-objective scheduling policy for allocating batches of serverless functions in the edge-cloud continuum. The scheduling policy minimizes energy consumption, makespan, resource usage and data transfers as compared to standard greedy algorithms. \\ \hline
Energy-efficient virtual machine scheduling in IaaS cloud environment using energy-aware green-particle swarm optimization ~\cite{ajmera2023energy} & The paper introduces \textbf{GPSO}, an energy-aware VM scheduling algorithm that minimizes power consumption and SLA violations in heterogeneous cloud data centers by prioritizing the use of the least active servers. \\ \hline

\multicolumn{2}{|c|}{\textbf{Energy Consumption Metrics Observation}} \\ \hline
Kepler: A Framework to Calculate the Energy Consumption of Containerized Applications†~\cite{kepler} & \textbf{Kepler (Kubernetes-based Efficient Power Level Exporter)} estimates real-time power consumption at the process, container, and pod levels by leveraging hardware counters and system power metrics. It uses a power model to predict workload energy consumption. \\ \hline
Container-level Energy Observability in Kubernetes Clusters ~\cite{pijnacker2025container} & This paper presents \textbf{KubeWatt}, an enhanced alternative to Kepler for energy measurement in Kubernetes clusters which is capable of accurately estimating power consumption at both node and container levels.  \\ \hline
\end{tabularx}
\vspace{-1ex}
\begin{flushleft}
\scriptsize † Industry paper
\end{flushleft}
\end{table}

\begin{table}[htbp]
\scriptsize
\caption{Summary of Carbon-Aware Scheduling Techniques in Kubernetes and Cloud Environments}
\noindent\textbf{\large Category: Carbon Awareness} \par\vspace{1ex}
\begin{tabularx}{\textwidth}{|p{4.5cm}|X|}
\hline
\textbf{Title} & \textbf{Key Proposal} \\ \hline

\multicolumn{2}{|c|}{\textbf{Fair Resource Allocation Strategies}} \\ \hline
Reducing Cloud Expenditures and Carbon Emissions via Virtual Machine Migration and Downsizing ~\cite{huang2023reducing} & Carbon intensity is reduced by algorithms that observe virtual machine resource waste to avoid idle or under-loaded virtual machines. \\ \hline
A Low Carbon Kubernetes Scheduler ~\cite{James2019ALC} & This scheduler calculates the cost of \textbf{relocating a data center} to assess if it should be re-dispatched to reduce carbon emissions. \\ \hline
Dominant Resource Fairness: Fair Allocation of Multiple Resource Types ~\cite{ghodsi2011dominant} & \textbf{Multi-resource max-min fairness} ensures that no node carries too many tasks, reducing excess emissions. \\ \hline
Ant Colony Algorithm for Multi-Objective Optimization of Container-Based Micro-Service Scheduling ~\cite{8744199} & \textbf{Ant colony algorithm} provides a load balancing scheduling algorithm for micro-services to reduce transmission overload and carbon emissions. \\ \hline

\multicolumn{2}{|c|}{\textbf{AI‑Driven and Deep Reinforcement Learning Schedulers}} \\ \hline
\textbf{Smart-Kube}: Energy-Aware and Fair Kubernetes Job Scheduler Using Deep Reinforcement Learning ~\cite{ghafouri2023smart} & \textbf{Smart-Kube} utilizes deep reinforcement learning to balance resource allocation and minimize energy consumption, ensuring efficient node utilization while maintaining low energy use. \\ \hline
DL2: A Deep Learning-Driven Scheduler for Deep Learning Clusters ~\cite{peng2021dl2} & \textbf{Deep learning clusters} are trained to dynamically and rationally adjust the resources allocated to a task, which is significantly more efficient than a fair scheduling program. \\ \hline
MAIZX: A Carbon-Aware Framework for Optimizing Cloud Computing Emissions ~\cite{MAIZX} & \textbf{MAIZX} is a dynamic, agent-driven framework that minimizes carbon emissions from cloud computing by optimizing workload allocation based on real-time carbon and energy metrics. \\ \hline
Greenflow: A carbon-efficient scheduler for deep learning workloads ~\cite{gu2024greenflow} & \textbf{GreenFlow} is a carbon-efficient GPU cluster scheduler that dynamically adjusts job configurations and resource allocations based on performance models and grid carbon intensity, minimizing deep learning training time within carbon budget constraints. \\ \hline

\multicolumn{2}{|c|}{\textbf{Serverless, Edge, and Federated Solutions}} \\ \hline
GreenCourier: Carbon-Aware Scheduling for Serverless Functions ~\cite{chadha2023greencourier} & \textbf{GreenCourier} schedules serverless functions based on regional carbon efficiencies, significantly reducing emissions per invocation by leveraging real-time carbon data from Watt-time and the Carbon aware SDK. \\ \hline
CAFE: Carbon-Aware Federated Learning in Geographically Distributed Data Centers ~\cite{cafe} & \textbf{CAFE} is scheduler uses federated learning to optimizes carbon emissions while ensuring data privacy by keeping data local. It uses the Lyapunov function to address prediction instability and environmental impacts. \\ \hline
CEFL: Carbon-Efficient Federated Learning ~\cite{cefl} & \textbf{CEFL} leverages machine learning training on edge devices shifts the focus from resource efficiency to resource cost, significantly reducing emissions without impacting training time. \\ \hline
Implementation and Benchmarking of Kubernetes Horizontal Pod Auto-scaling Method to Event-Driven Messaging System ~\cite{KEDA} & \textbf{KEDA} is an event-driven auto-scaling component for Kubernetes that optimizes resource utilization by dynamically scaling containers based on events. \\ \hline
CASA: A Framework for SLO- and Carbon-Aware Auto scaling and Scheduling in Serverless Cloud Computing ~\cite{qi2024casa} & \textbf{CASA} is a carbon and service level
objectives(SLOs) aware scheduling framework for serverless platforms that reduces emissions and lowers the risk of SLO violations by intelligently balancing container warm-up and auto-scaling decisions. \\ \hline
GreenWhisk: Emission-Aware Computing for Serverless Platform ~\cite{serenari2024greenwhisk} & \textbf{GreenWhisk} is a serverless computing platform that reduces carbon emissions while maintaining performance by scheduling jobs based on real-time energy and carbon intensity data. \\ \hline

\multicolumn{2}{|c|}{\textbf{Carbon Emission Index Priority Scheduling Frameworks}} \\ \hline
Carbon Emission-Aware Job Scheduling for Kubernetes Deployments ~\cite{piontek2024carbon} & Introduces a CO$_2$-aware workload scheduling algorithm that shifts non-critical jobs in time using historical energy data to reduce emissions. \\ \hline
CASPER: Carbon-Aware Scheduling and Provisioning for Distributed Web Services ~\cite{Souza_2023} & This paper proposes a Kubernetes-deployable architecture named \textbf{CASPER} to reduce the carbon footprint of geographically distributed web services through spatial shifting while minimizing performance costs. \\ \hline
PCAPS: Precedence- and Carbon-Aware DAG Scheduling ~\cite{lechowicz2025pcaps} & \textbf{PCAPS} is a precedence and carbon aware scheduler for data processing jobs that reduces emissions by up to 32.9\% while maintaining performance by avoiding bottlenecks in task dependencies. \\ \hline
Operating Cloud Applications Under a Carbon Budget ~\cite{kreutz2025budget} & This work introduces a carbon-aware microservice deployment approach that dynamically selects configurations to maximize user experience and revenue while adhering to \textbf{hourly carbon budgets}. \\ \hline
U-DUCT: Uncertainty-aware Dynamic Unified Carbon Modeling Tool for Datacenter Scheduling ~\cite{guan2024u} & \textbf{U-DUCT} is a comprehensive runtime-aware carbon modeling tool that captures hardware and software uncertainties in computing, storage, and networking to improve carbon emissions assessment and mitigation measures in data centers. \\ \hline
A Green Cloud-Based Framework for Energy-Efficient Task Scheduling Using Carbon Intensity Data for Heterogeneous Cloud Servers ~\cite{beena2025green} & This paper proposes a scalable scheduling framework based on Kubernetes that integrates real-time carbon intensity data to dynamically optimize \textbf{high-energy workloads} across cloud platforms, significantly improving energy efficiency and supporting sustainable development goals. \\ \hline
Carbon-Aware Temporal Data Transfer Scheduling Across Cloud Datacenters ~\cite{rodrigues2025carbon} & \textbf{LinTS} is a carbon-aware time-based data transmission scheduler that significantly reduces emissions from cross-data center communications by aligning transmissions with low-carbon periods while meeting all task deadlines. \\ \hline
\end{tabularx}
\vspace{-1ex}
\begin{flushleft}
\scriptsize † Industry paper
\end{flushleft}
\end{table}

Collectively, both hardware and software optimization strategies are central to enabling scalable, environmentally sustainable container scheduling in cloud-native systems. These research efforts represent critical steps toward carbon-aware and energy-efficient computing infrastructure.

\section{Detailed Hardware Optimized Algorithm}
\subsection{Energy Efficiency Sustainability Goal}
\subsubsection{Energy Consumption Priority Scheduler}
\textbf{\newline Holistic Scheduling}
Monitor the hardware response of each server in the data center with predictive capabilities to assess the impact of each new incoming container in the system so that containers can be assigned to nodes more optimally to reduce power consumption, improve performance, or balance between the two.

Server hardware behavior modeling translates multiple influences such as ambient temperature, power consumption, and internally recorded server metrics into a multivariate polynomial equation. This can be considered an optimization problem and can be applied to a variety of different types of data centers.

In addition to collecting hardware metrics, a large number of software metrics need to be collected from the containers running in the distributed cluster. Since collecting software metrics also incurs significant overhead and the amount of data to be analyzed can be quite large, it is appropriate to collect software metrics at a frequency that provides an acceptable level of accuracy, but not so frequently as to cause excessive storage or analysis overhead.

Through large-scale testing, the scheduler using a combination of hardware and software metrics scheduling methodology consumes 10\%-20\% less power with essentially the same performance as Kubernetes~\cite{townend2019improving}.

\textbf{HEATS Scheduling Strategy}
Each cloud provider has different CPU architectures, GPUs, FGPAs, ASICs, or whether the processor frequency can be dynamically increased or decreased. And in order to accommodate different resource requirements, tasks can be migrated from one cloud provider machine to another to meet the demand. The HEATS container scheduling strategy utilizes the underlying hardware resource model to place the tasks on the most appropriate currently available hardware resource nodes.

Gather resource requirements before task execution, use Heapster to collect hardware state, and model hardware nodes with different architectural runtimes and energy consumption. Periodically, the energy consumption to performance ratio of each node is calculated as weights, and tasks are assigned to the highest scoring node. Each computer architecture can reduce energy consumption at the expense of performance, but in different ratios. Testing shows that this approach reduces energy consumption by 1.5\% compared to Kubernetes' scheduling strategy without compromising performance. If energy consumption is prioritized, it can be reduced by up to 7.1\%~\cite{rocha2019heats}.

\textbf{PEAKS: Power-Efficiency-Aware Kubernetes Scheduler}
The core idea of \textbf{PEAKS (Power-Efficiency-Aware Kubernetes Scheduler)}~\cite{souza2024peaks} is to extend Kubernetes scheduling by incorporating node-level power efficiency metrics to improve energy utilization without degrading performance. PEAKS leverages the Kepler project to collect power-related telemetry (such as CPU power, memory power, and energy per instruction) and assigns each node a power efficiency score. During scheduling, it filters available nodes and ranks them not only by resource availability but also by their power efficiency scores, preferring nodes that can perform more work per watt. This allows Kubernetes to make energy-aware scheduling decisions, reducing overall power consumption and carbon footprint while maintaining Quality of Services(QoS).

\textbf{Real-Time Node’s Power-Aware Kubernetes Scheduler(NPAKS) in a Cloud Environment}
This paper proposes \textbf{NPAKS}~\cite{kumari2025npaks}, an energy-aware scheduler for Kubernetes that enhances energy efficiency by incorporating real-time energy consumption metrics of cluster nodes into scheduling decisions. Unlike the default Kubernetes scheduler, which primarily considers resources such as CPU and memory, NPAKS utilizes real-time telemetry data collected by Prometheus and Kepler to identify and prioritize nodes with lower energy consumption. By deploying Pods on energy-efficient nodes while meeting service level agreements (SLAs), NPAKS aims to reduce overall energy consumption in cloud environments. This method is implemented as a custom plugin within the Kubernetes scheduling framework and has demonstrated quantifiable energy savings in testbed experiments.

\textbf{Fine-Grained Heterogeneous Execution Framework with Energy Aware Scheduling}
This paper proposes an energy-aware scheduling framework based on Kubernetes, specifically optimized for fine-grained task execution in heterogeneous computing environments involving CPUs, GPUs, and FPGAs. By integrating with technologies such as NVIDIA's Multi-Process Service (MPS), the framework enables more efficient utilization of hardware accelerators, facilitating the collaborative execution of multiple lightweight tasks on shared resources. By combining energy-aware scheduling with hardware-specific scheduling, the framework improves workload completion time and reduces energy consumption for CPU and GPU workloads. This approach enhances performance and sustainability in serverless and data-intensive computing environments~\cite{rattihalli2023fine}.

\subsubsection{Data Center Cost Management Strategies}
\textbf{\newline Virtual Data Center Deployment Model based on the Green Cloud Computing}
The technology in this paper is based on virtualization and reduces energy consumption and carbon footprints in cloud computing by introducing a green cloud management framework. Unlike traditional virtual machine-based services, this framework uses a virtual data center (VDC) model to group virtual machines according to network communication patterns. These VDCs are then strategically deployed in green data centers to optimize energy efficiency, minimize environmental impact, and maximize service provider revenue. While the framework predates container orchestration platforms like Kubernetes, its principles can be extended to inform energy-aware scheduling strategies in modern Kubernetes-based infrastructures~\cite{xu2014virtual}.

\textbf{Carbon-Efficient Virtual Machine Placement in Cloud Datacenters over Optical Networks}
This paper introduces an integer linear programming (ILP) model which seeks to minimize carbon emissions by determining the optimal deployment locations for virtual machines (VMs) across geographically distributed data centers. The model takes into account key constraints such as latency, bandwidth, and resource capacity between data centers, while also incorporating each data center’s power usage effectiveness (PUE) and carbon efficiency (CE) metrics to guide deployment decisions. Simulation results demonstrate that migrating VMs to more environmentally friendly regions can achieve significant emission reductions without violating service level requirements. Although the model was initially designed for VMs similar to Virtual Data Center, the proposed carbon-aware deployment strategy can be adapted to modern Kubernetes-based environments by integrating carbon intensity and latency metrics into container scheduling strategies, thereby enabling sustainable orchestration of containerized workloads~\cite{zhang2023carbon}.

\subsection{Carbon Awareness Sustainability Goal}
\subsubsection{Multi-Criteria Optimization Methods}
\textbf{Static and Dynamic Temperature-Aware Scheduling for Multiprocessor SoCs}
This paper proposes a fundamental method for temperature-aware task scheduling in multi-processor system-on-chip (MPSoC) systems, aiming to reduce thermal hotspots, spatial temperature gradients, and thermal cycling, which can degrade system reliability and increase cooling and leakage costs~\cite{coskun2008static}. The method introduces both ILP-based static scheduling strategies and dynamic operating system-level scheduling strategies, significantly improving thermal behavior while maintaining performance. The static ILP model achieves performance improvements by optimizing energy consumption, balance, and thermal uniformity, while the dynamic scheduler adaptively reduces high-intensity thermal effects by minimizing overhead. Although this is an early research result, it still has significant influence and practical significance in energy efficiency perception and thermal perception scheduling research. This is because it lays the foundation for integrating thermal constraints into system-level scheduling strategies, a concept that is becoming increasingly important in modern multi-core and edge computing platforms.

\textbf{KCSS Scheduling Algorithm}  
The Kubernetes Container Scheduling Strategy (KCSS) is a multi-criteria scheduler designed to minimize scheduling and execution time as well as overall power consumption~\cite{menouer2021kcss}. Built on Kubernetes and inspired by tools like Docker SwarmKit and Apache Mesos, \textbf{KCSS} balances user time constraints with cloud provider energy efficiency by selecting optimal nodes using the TOPSIS algorithm.

\textbf{KCSS} currently considers six criteria:
\begin{enumerate}
    \item Maximize CPU usage: Deploy tasks to nodes with high CPU utilization when future requirements are unknown.
    \item Maximize memory usage: Select nodes with high memory usage under similar uncertainty.
    \item Maximize disk usage: Prioritize nodes with higher disk usage to consolidate workloads.
    \item Minimize power consumption: Use CloudSim Plus to estimate and select nodes with lower energy demand.
    \item Minimize running containers: Distribute load evenly for resilience.
    \item Minimize deployment time: Favor nodes with cached container images to reduce network overhead.
\end{enumerate}

Experiments demonstrate that \textbf{KCSS} outperforms the default Kubernetes scheduler and other standard strategies in reducing task latency, power consumption, and container wait times for large-scale workloads.

\textbf{Carbon-Aware Online Control}

The carbon-aware control framework manages cloud services distributed across various geographical locations in a way that minimizes carbon emissions. The framework exploits the spatial and temporal variabilities in the carbon footprint of electricity across different regions~\cite{zhou2015carbon}. The framework focuses on three key areas:

\begin{itemize}
    \item \textbf{Geographical Load Balancing:} Distributes workloads across different data centers based on their carbon efficiency.
    
    \item \textbf{Capacity Right-Sizing:} Adjusts the number of active servers in each data center to match the current demand efficiently.
    
    \item \textbf{Server Speed Scaling:} Dynamically adjusts server speeds to balance performance and energy consumption.
\end{itemize}

\textbf{Lyapunov Function Construction} 
A Lyapunov function is constructed to represent the system’s state and stability. This function helps in monitoring the system's performance and ensuring that it remains within desired operational bounds.

\textbf{Drift-Plus-Penalty Minimization} 
The system aims to minimize the Lyapunov drift-plus-penalty, which ensures that the system remains stable while optimizing the desired objectives. The drift represents the change in the Lyapunov function, while the penalty represents the cost or emission metrics that need to be minimized.

\textbf{Online Decision Making} 
Based on real-time data, the framework makes online decisions regarding load balancing, capacity right-sizing, and server speed scaling. These decisions are made iteratively, ensuring that the system adapts to changing conditions and continues to optimize performance.

\subsubsection{Temporal and Spatial Carbon Shifting Systems}
\textbf{Carbon Aware Computing}
While digital transformation can reduce emissions from physical activities, software itself contributes to carbon output through energy consumption. Carbon-aware computing aims to reduce this impact by aligning workloads with cleaner energy availability, primarily via time shifting~\cite{Microsoft2023}.

For instance, in solar-rich regions, carbon intensity is lower midday. Shifting short, high-load tasks to such windows reduces emissions. Key criteria include intensity, duration, capacity, and startup time. The Software Carbon Intensity (SCI) metric quantifies emissions, calculated as:

\[
\text{SCI} = \frac{(\text{Energy Use} \times \text{Carbon Intensity}) + \text{Hidden Emissions}}{\text{Software Size}}
\]

Optimization involves (1) measuring intensity over time/location, (2) predicting optimal windows, (3) comparing with actual usage, and (4) evaluating SCI reduction.

\textbf{CarbonScaler: Leveraging Cloud Workload Elasticity for Optimizing Carbon-Efficiency}
The core idea of this paper is to reduce the carbon footprint of cloud batch processing workloads by introducing carbon scaling, a dynamic resource allocation strategy. Carbon scaling is a strategy that dynamically adjusts server utilization rate based on the real-time carbon intensity of the power grid~\cite{hanafy2023carbonscaler}. Unlike traditional pause-resume methods, which can significantly delay task completion times, carbon scaling leverages the elasticity of batch processing tasks to minimize emissions while maintaining performance. The authors propose a greedy marginal resource allocation algorithm and implement it as a Kubernetes-based prototype system called CarbonScaler, which leverages Kubernetes' native auto scaling capabilities. By adjusting computing resources based on carbon signals, CarbonScaler achieves significant emissions reductions of up to 51\% compared to carbon-agnostic execution. This demonstrates that Kubernetes can serve as an effective platform for carbon-conscious workload orchestration.

\textbf{Caspian: Carbon-Aware Multi-Cluster Scheduling for Cloud-Native Workloads}
\textbf{Caspian}, developed by IBM, is a carbon-aware scheduler for multi-cluster Kubernetes environments that minimizes the carbon footprint of containerized workloads by leveraging spatial and temporal variations in electricity carbon intensity. It dynamically schedules tasks across geographically distributed clusters and optimal time windows based on real-time or predictive carbon data, while maintaining Quality of Service (QoS) and service level objectives (SLOs). \textbf{Caspian} integrates with the Multi Cluster App Dispatcher (MCAD) and the Kubernetes scheduling framework, using optimization algorithms and external carbon intensity APIs to inform intelligent workload placement and timing decisions. Experimental results show that \textbf{Caspian} can reduce carbon emissions by approximately 33\%, and 98\% of workloads are completed on time without compromising performance, achieving sustainability~\cite{ibm2024caspian}.

\textbf{S.C.A.L.E.: Scheduler for Carbon-Aware Load Execution at ING }
\textbf{SCALE} is a practical carbon-aware scheduling framework developed by ING, designed to reduce greenhouse gas emissions from data centers. \textbf{SCALE} targets resource-intensive batch pipelines, intelligently scheduling tasks during periods when renewable energy supply is abundant and grid carbon intensity is low. The framework comprises three core modules: a module for predicting task execution times, a module for forecasting green energy supply and grid demand, and a module for integrating with existing data pipelines. By shifting flexible batch workloads to “green energy periods,” \textbf{SCALE} is expected to achieve a 20\% reduction in carbon emissions while meeting task deadlines. The paper also highlights that effects may vary due to seasonality and data arrival patterns, but overall, it demonstrates the feasibility and benefits of carbon-aware scheduling in large Kubernetes-based data systems in real-world scenarios.

\subsubsection{Carbon-Aware Workload Shifting}
\textbf{Microsoft Carbon-Aware Computing}
Microsoft's “Carbon-Aware Computing White Paper” proposes a framework for reducing the carbon footprint of software by enabling systems to be time and location aware~\cite{Microsoft2023}. The core concept involves collaborating with the Green Software Foundation and WattTime to obtain real-time carbon emission intensity data from various regions, and then shifting workloads to cleaner time slots and regions within the power grid, thereby minimizing emissions without compromising performance. To this end, Microsoft has developed an open-source Carbon-Aware Software Development Kit (Carbon-Aware SDK), which enables developers to make intelligent scheduling decisions based on real-time or predictive carbon data. The white paper also introduces the Software Carbon Intensity (SCI) specification, a standardized metric for quantifying the carbon impact of software. Through a collaboration with UBS, Microsoft demonstrates that carbon-aware workload shifting can achieve significant emission reductions, highlighting the potential of embedding sustainability directly into software architecture.

\textbf{Intel Telemetry Aware Scheduling}
Intel adjusts the load through external telemetry, such as CPU and power usage. By setting a threshold for Telemetry Aware Scheduling (TAS), Kubernetes' Horizontal Pod Autoscaler(HPA) is triggered when a pod's threshold is elevated, then reducing emissions at the physical level by creating more pods~\cite{intel2024tas}.

\section{Detailed Software Optimized Algorithm}
\subsection{Energy Efficiency Sustainability Goal}
\subsubsection{Machine Learning and Deep Learning Based Scheduling Approaches}

\textbf{\newline An RL-Based Model for Optimized Kubernetes Scheduling:}
The core idea of this paper is to develop \textbf{RLKube}, a custom Kubernetes scheduler plugin that uses reinforcement learning (RL) to improve task scheduling efficiency in Kubernetes (K8s) clusters~\cite{rothman2023rl}. \textbf{RLKube} aims to maximize resource utilization, improve Pod throughput, and enhance energy efficiency, thereby overcoming the limitations of default K8s scheduling strategies, such as Least-allocated and Most-allocated. By adopting a Dual-Depth Q-Network (DDQN) with Priority Experience Replay (PER), \textbf{RLKube} is trained to make scheduling decisions based on multiple optimization objectives. Multiple reward functions are defined to target specific objectives, such as energy savings, fairness, and performance.

\textbf{RLKube} is directly integrated into the Kubernetes scheduling pipeline as a scoring plugin, with its decisions based on real-time cluster metrics collected via Prometheus and Node Exporter. The system has been tested on both real and synthetic workloads, showing significant improvements in throughput and machine utilization compared to default strategies.

This work is significant for Kubernetes-based infrastructure as it demonstrates how to effectively leverage machine learning (particularly RL) to design intelligent and adaptive scheduling strategies, making it an important reference for sustainable, high-performance, and dynamic orchestration in modern cloud and edge environments.

\textbf{Optimizing Energy Consumption of Kubernetes Clusters with Deep Reinforcement Learning}
This paper proposes a method based on \textbf{deep reinforcement learning (DRL)} aimed at reducing the energy consumption of Kubernetes clusters. By training a Proximal Policy Optimization (PPO) agent using a custom neural network, the system can learn an optimized pod placement strategy that balances the needs of newly created and running pods. To enforce the required resource allocation in the Kubernetes environment, researchers developed a Kubernetes Operator based on the Operator Framework. This Operator introduces a custom resource definition (CRD) named PodPlacement, enabling real-time integration with the control plane. Evaluated in heterogeneous tests simulating edge cloud environments, the DRL-based scheduler achieved up to 24\% energy savings compared to the default Kubernetes scheduler, with particularly significant results under unsaturated cluster conditions. This method achieves a balance between sustainability and performance without compromising Pod execution success rates~\cite{espinosa2024optimizing}.

\textbf{EETS: An energy-efficient task scheduler in cloud computing based on improved DQN algorithm}
EETS (Energy-Efficient Task Scheduler) is a deep reinforcement learning based task scheduling framework designed to minimize energy consumption and task response time in cloud computing environments. EETS is based on an improved Deep Q-Network (DQN) architecture by integrating \textbf{Double DQN and Dueling DQN (D3QN)} to address overestimation bias and improve value stability. It also incorporates Prioritized Experience Replay (PER) to improve sample learning efficiency during training. The scheduler models the environment as a batch-based Markov Decision Process (MDP) where tasks are assigned to virtual machines (VMs) based on their attributes and current VM states. Its reward function balances energy use, waiting time and execution time using tunable weights and energy consumption is estimated using a linear model based on CPU utilization. The evaluation done on Alibaba’s Cluster Trace with 20 heterogeneous VMs shows that EETS outperforms various heuristic, meta-heuristic and DRL baselines across different workload sizes. While the current simulation is on independent tasks, the authors propose extending the framework to support Directed Acyclic Graphs (DAGs) for scheduling interdependent tasks~\cite{hou2024eets}.

\textbf{An energy efficient RL based workflow scheduling in cloud computing}
This paper presents an energy-efficient workflow scheduling framework for cloud computing that leverages \textbf{Reinforcement Learning (RL)} to optimize multiple objectives, including energy consumption, execution time, and cost. The framework also integrates security through the \textbf{X-NOR Whirlpool hashing algorithm} to ensure that only legitimate access is granted to the cloud. For optimization and monitoring, the framework integrates several advanced techniques: \textbf{LWMA-Sea Lion Optimization} for parameter minimization, \textbf{Jordan Normal Form-Deep Kronecker Neural Network(JNF-DKNN)} for resource monitoring, and \textbf{Fuzzy Self-Defense Algorithm(CD-FSDA)} for selecting efficient virtual machines to schedule tasks. Evaluation on standard datasets reveals that the proposed \textbf{LJC(LWMA + JNF-DKNN + CD-FSDA)} framework exhibits superior performance as well as higher security compared to existing algorithms~\cite{reddy2023energy}.

\textbf{HunterPlus: AI based energy-efficient task scheduling for cloud-fog computing environments}
This paper applies deep learning to cloud-fog task scheduling to improve the energy efficiency by building on the existing \textbf{HUNTER} model, an AI based holistic resource management technique for sustainable cloud computing. HUNTER utilizes a Gated Graph Convolution Network (GGCN) as a surrogate model to approximate the Quality of Service (QoS) for a given system state and generate optimal scheduling decisions. HunterPlus, an extension of the HUNTER model, explores two enhancements. First, the original GGCN’s gated recurrent unit (GRU) is extended with a \textbf{bidirectional GRU} to process the input graph sequences both in forward and backward direction. This added temporal context leads to more stable and consistent resource allocation decisions with improvements in energy savings compared to uni-directional GGCN. The second contribution is a \textbf{CNN-based surrogate scheduler} in which resource-task allocation states are encoded as 2D arrays ("images") and processed with a Convolutional Neural Network. This surrogate model learns the mapping between host-task matrices and energy outcomes, enabling gradient-based optimization for real-time scheduling. This enhancement achieved at least 17\% reduction in energy consumption per task and 10.4\% improvement in job completion rate compared to both GGCN variants~\cite{tuli2022hunter}.

\textbf{Energy-Efficient Cloud Computing Through Reinforcement Learning-Based Workload Scheduling}
To address load balancing, the model computes a VM utilization variance metric and triggers task migration when the imbalance exceeds a defined threshold. Tasks for migration are selected based on the complexity-to-resource ratio to ensure that high-priority jobs are scheduled efficiently. The authors simulate dynamic workload scenarios and assess model performance across parameters like latency, throughput, resource utilization, load balancing and QoS. Based on evaluation output, the model reduce latency to 15 ms and throughput up to 500 tasks/sec with 92\% efficiency in load balancing, 95\% resource usage and 97\% QoS~\cite{malipatil2025energy}.

\textbf{Opportunistic Energy-Aware Scheduling for Container Orchestration Platforms Using Graph Neural Networks}
This paper proposes an opportunistic energy-aware container scheduling framework for Kubernetes that uses a Graph Neural Network (GNN) to predict the power consumption of container workloads based on telemetry data and application signatures. A novel graph model is developed to represent host machines using a heterogeneous graph structure composed of nodes representing various hardware resources (CPU, memory, disk, network, GPU, FPGA, SmartNIC, temperature sensors etc) and edges encoding physical or semantic relationships between them. Telemetry data is collected every second from Prometheus-based exporters (cAdvisor, NodeExporter, Nvidia DCGM, HPE iLO, Xilinx xbutil), aggregated into time frames and used to generate feature vectors for each node. The application’s resource usage signature is captured via container-level metrics for a subset of sensors and a merge function is defined to simulate \emph{what-if} scheduling scenarios by combining this signature with the current host graph. The authors use a Heterogeneous Graph Transformer stacked with pooling layers and linear transformations to predict min, max and average power consumption over a time frame. The model achieves a mean RMSE of 7.5\% on profiling data collected from AMD 7443 dual-socket CPUs and can distinguish between similar hosts with different idle power due to hardware accelerators like GPUs~\cite{raith2024opportunistic}.

\textbf{Energy-efficient DAG scheduling with DVFS for cloud data centers}
This study proposes E2DSched, a reinforcement learning-based hierarchical scheduler for \textbf{online Directed Acyclic Graph} (DAG) job scheduling in heterogeneous cloud data centers, focusing on joint optimization of energy consumption and quality of service (QoS). E2DSched integrates \textbf{Dynamic Voltage and Frequency Scaling} (DVFS) into its decision-making and decomposes the scheduling process into three distinct layers: \emph{task selection}, \emph{server selection} and \emph{frequency control}. Each layer is managed by a separate PPO-based reinforcement learning agent which enables fine-grained decisions while avoiding the convergence issues associated with large action spaces. A key contribution is the development of a lightweight and accurate energy consumption model that predicts server power usage based on CPU frequency and load and is validated with around 5\% error on Intel Xeon servers. E2DSched is evaluated using the BitBrain dataset and scientific workflows across cluster sizes ranging from 10 to 100 nodes. Results show that E2DSched can reduce energy consumption in heterogeneous clusters without signifcantly compromising quality of service compared to traditional methods~\cite{yang2024energy}.

\subsubsection{Heuristics, Metaheuristics, and Analytical Approaches}
\textbf{KEIDS: K8S Energy and Interference Driven Scheduler}
KEIDS(Kubernetes-Based Energy and Interference Driven Scheduler) is based on three core principles

1. Minimize carbon intensity by optimizing the use of green energy.

2. Minimize IoT interventions between devices while achieving superior performance.

3. Minimize software and hardware consumption.

By using these principles as a multi-objective optimization problem, the energy utilization of edge cloud nodes is improved. This enables faster scheduling of applications to available nodes while reducing interference from other IoT devices. Ultimately, this approach ensures optimal performance for end users~\cite{kaur2019keids}.

\textbf{Energy-aware Scheduling Algorithm for Microservices in Kubernetes Clouds}
This paper addresses the inefficiency of Kubernetes' default scheduling in distributed microservices by introducing an energy-aware scheduling algorithm that considers service level agreement (SLA) constraints and cross-service communication patterns. The method quantifies communication frequency based on network traffic and prioritizes Pods according to resource consumption. To improve placement efficiency, the authors integrate an improved swallow search algorithm (ISSA), which optimizes Pod packaging by combining communication intensity and resource requirements. The method aims to reduce CPU overhead and energy waste caused by heartbeat mechanisms and Pod-to-Pod communication. Experimental evaluations in cloud environments demonstrate that the method can reduce total cluster energy consumption by at least 5\% while maintaining SLA compliance~\cite{rao2024sla}.

\textbf{Analyzing Energy-Efficient and Kubernetes-Based Auto-scaling of Microservices Using Probabilistic Model Checking}
This paper proposes an energy-efficient method for analyzing automatic scaling strategies of applications in cloud environments based on microservices. The authors propose multiple Markov decision process (MDP) models in the Kubernetes Horizontal Pod Autoscaler (HPA) to capture different constraints and scaling behaviors. They use probabilistic model checking (PMC) to evaluate the energy consumption and SLA violations caused by these automatic scaling strategies. These models are encoded using action-based boundaries (BBA) and state-based boundaries (BBS) techniques, enabling formal verification and sensitivity analysis of scaling decisions. The research findings indicate that combining latency with energy consumption metrics in scaling strategies yields optimal energy efficiency results. This framework provides cloud engineers with a method to evaluate and select auto-scaling strategies that minimize energy consumption while meeting service objectives~\cite{Jawaddi2025}.

\textbf{GreenPod: Energy-Optimized Scheduling for AIoT Workloads Using TOPSIS}
This paper introduces \textbf{Greenpod}, a multi-criteria, energy-aware Kubernetes scheduler designed specifically for Artificial Intelligence of Things(AIoT) workloads in heterogeneous cloud edge environments. The scheduler uses the TOPSIS decision-making framework to evaluate factors such as execution time, energy consumption, CPU core count, memory availability, and load balancing to determine the optimal Pod placement. When tested on actual Google Kubernetes Engine (GKE) clusters, \textbf{Greenpod} achieved up to 39.1\% energy savings compared to the default Kubernetes scheduler, particularly in workloads with moderate complexity. Test results demonstrate that \textbf{Greenpod} can effectively balance sustainability and performance with minimal scheduling latency, making it suitable for a wide range of resource-sensitive applications~\cite{pradeep2025energy}.

\textbf{3GC: A deadline-aware and energy-efficient resource allocation scheme for serverless edge computing}
The core idea of this paper is to improve the energy efficiency and cost-effectiveness of performing tasks in serverless edge environments. The proposed approach named \textbf{Go-Green-Go-Cheap (\textbf{3GC})} introduces a deadline-aware real-time resource allocation framework that utilizes Dynamic Voltage and Frequency Scaling (DVFS) to balance latency and energy consumption. The framework is designed for Kubernetes-based platforms and aims to minimize execution time and energy costs while meeting latency requirements. Evaluation results show that \textbf{3GC} reduces operational costs by 39.35\% to 69.43\%, outperforming existing allocation strategies without violating functional deadlines~\cite{bellal20253gc}.

\textbf{FOA-Energy: A Multi-objective Energy-Aware Scheduling Policy for Serverless-based Edge Cloud Continuum}
This paper presents \textbf{FOA-energy (Function Orchestration Algorithm, version energy)}, a \emph{two-level multi-objective scheduling policy} designed for serverless computing in the heterogeneous edge-cloud continuum and uses Kubernetes as the orchestration backbone. FOA-energy addresses challenges in serverless environments including energy consumption, cold start delays, makespan, data transfer overhead and platform heterogeneity. The scheduling policy simultaneously optimizes three objectives: minimizing energy consumption, makespan and data transfers. It operates across two levels of the continuum: the global level (across clusters) and the local level (within clusters). At the global level, FOA-Energy solves a linear programming (LP) formulation to allocate functions and container environments to clusters while minimizing energy cost and data transfers under a makespan constraint. This fractional solution is then transformed into an integral schedule using a minimum-cost integral matching method inspired by the \emph{Shmoys-Tardos algorithm} ~\cite{shmoys1993approximation}. At the local level, the policy uses one of four traditional strategies including First-fit, First Come First Serve(FCFS), Smallest-first and  Largest-first to assign functions to individual machines while also considering container layer reuse to minimize image downloads and reduce cold start delays. FOA-energy is evaluated using 1,350 experiments across bare-metal and simulated environments showing that it outperforms a Kubernetes-based baseline by up to three orders of magnitude in energy, makespan and data transfer~\cite{da2025foa}.

\textbf{Energy-efficient virtual machine scheduling in IaaS cloud environment using energy-aware green-particle swarm optimization}
This paper presents a \textbf{Green Particle Swarm Optimization (GPSO)} algorithm for optimizing virtual machine (VM) consolidation in cloud data centers consisting of heterogeneous servers.The \textbf{GPSO} algorithm recognizes the trade-offs between reducing energy consumption and maintaining service level agreement (SLA) compliance, and thus identifies energy-efficient “green” servers (green particles) and schedules virtual machines to minimize the number of active servers. This reduces power consumption while controlling SLA violations. When implemented in CloudSim, \textbf{GPSO} outperforms existing VM scheduling algorithms in terms of energy efficiency and quality of service~\cite{ajmera2023energy}.

\subsubsection{Energy Consumption Metrics Observation}
\textbf{\newline Kepler: A Framework to Estimate Energy Consumption}
Kepler (Kubernetes-based Efficient Power Level Exporter) is an open-source framework for estimating power consumption at the process, container, and pod levels in Kubernetes clusters~\cite{kepler}. It leverages hardware-level APIs (e.g., Intel RAPL, NVML) and eBPF to extract real-time system metrics.

Power is categorized into:
\begin{itemize}
    \item \textbf{Idle Power} – baseline consumption at rest
    \item \textbf{Dynamic Constant Power} – load-independent activity power
    \item \textbf{Dynamic Variable Power} – load-dependent power
\end{itemize}

Kepler includes three components:
\begin{itemize}
    \item \textbf{Metric Exporter} collects process-level data.
    \item \textbf{Model Server} trains regression models (e.g., XGBoost, SVR) to estimate power.
    \item \textbf{Inference Module} predicts consumption, aggregated using cgroups, and exports via Prometheus.
\end{itemize}

Compared to traditional aggregated models (MSE: 0.92), Kepler achieves a much lower MSE of 0.010. It enables three scheduling algorithms:
\textit{GNN}, \textit{GNN-Aware}, and \textit{GNN-Packing}, which use container signatures and host states to minimize energy use. GNN-based scheduling reduces energy consumption by 6.2\% without affecting workload makespan~\cite{raith2024opportunistic}.

\textbf{Container-level Energy Observability in Kubernetes Clusters}
This paper proposes KubeWatt, an energy observability tool for Kubernetes clusters that addresses key limitations in the existing CNCF project Kepler, which estimates container-level power consumption. While Kepler relies on utilization metrics and power models ( RAPL, NVML, Redfish) to estimate node and container power, it exhibits inaccuracies such as attributing idle power to non-running containers and misallocating dynamic power to undefined system processes. To evaluate this, the authors develop a controlled testbed using iDRAC and Redfish APIs on a Dell PowerEdge server for ground-truth node-level power data and Prometheus/cAdvisor for CPU and container metrics. Experimental results show Kepler has a container-level RMSE of 66.4W and underperforms during transitions like pod deletions. On the other hand, KubeWatt introduces a refined allocation model that separates static (idle) and dynamic power, where dynamic power is proportionally distributed among running containers based on their CPU usage. KubeWatt operates in three modes: base initialization (on idle clusters), bootstrap initialization (using linear regression on sub-50\% CPU utilization data) and estimation mode (for real-time monitoring). KubeWatt estimates static power within 0.2\% of ground-truth measurements and provides accurate dynamic power attribution across containers. Its design makes it a practical and reliable foundation for fine-grained energy and carbon-aware scheduling in Kubernetes environments.

\subsection{Carbon Awareness Efficiency Sustainability Goal}
\subsubsection{Fair Resource Allocation Strategies}
\textbf{\newline Reducing Cloud Expenditures and Carbon Emissions via Virtual Machine Migration and Downsizing}
Data centers have a large number of virtual machines running, consuming more and more energy and having a larger carbon footprint. If a balance can be struck between VM performance and carbon intensity, carbon intensity can be reduced by sacrificing the least amount of performance. Experiments have shown that carbon intensity can be effectively reduced by allowing some latency, reducing some bandwidth and increasing the capacity of green data centers~\cite{xu2014virtual}.

Data centers can also be virtualized like virtual machines and place virtual data centers in low-emission green data centers. However, it's crucial to consider external factors such as the costs associated with green data centers and electricity consumption. By carefully balancing these factors, providers can maximize benefits while simultaneously reducing carbon emissions.

By analyzing data from over 2.6 million virtual machines and their carbon intensity metrics, we can quantify the energy inefficiency of cloud computing. Implementing VM scheduling algorithms based on this analysis can effectively reduce carbon emissions and lower the costs associated with cloud computing operations.

Machine learning algorithms are utilized to predict the carbon intensity of tasks at different time intervals. Using these predictions, tasks can be dynamically rescheduled to minimize carbon intensity by following to predefined thresholds or scheduling tasks during periods of lowest average carbon intensity. This approach optimizes resource utilization and helps reduce the environmental impact of computing operations.

VM waste can be obtained by multiplying the cost by the VM's CPU unused rate.

waste = cost * (1 - CPU Utilization)

To reduce excessive waste or inefficiency in VMs, consider reducing the number of cores without impacting essential services. In addition, identify and shut down VMs with low CPU utilization to optimize resource usage and maintain operational efficiency~\cite{huang2023reducing}.

\textbf{A Low-Carbon Kubernetes Scheduler: Demand Side Management}  
Demand Side Management (DSM) enables users to monitor and adjust energy use to reduce peak demand, improve reliability, and lower costs~\cite{James2019ALC}. In a low-carbon scheduling context, task placement decisions are guided by comparing the emissions of local and remote data centers.

Let:
\begin{itemize}
    \item $EC_{A}, EC_{B}$: Compute energy at Data Centers A and B
    \item $I_{A}, I_{B}$: Carbon intensity at A and B
    \item $ER_{B}$: Deployment energy at B
    \item $EN_{AB}$: Transfer energy from A to B
    \item $I_{AB}$: Transfer carbon intensity
\end{itemize}

A migration to Data Center B is preferred if:
\begin{equation}
    EC_{A}I_{A} > EC_{B}I_{B} + ER_{B}I_{B} + EN_{AB}I_{AB}
\end{equation}

\textbf{The Heliotropic Scheduler (‘follow-the-sun”)}
The “follow the sun” model addresses the challenges posed by the intermittency of solar energy, which cannot consistently meet electricity standards in one location throughout the day. This model minimizes emissions by strategically shifting applications to areas with higher solar availability as the sun moves across the globe~\cite{James2019ALC}.

\textbf{Dominant Resource Fairness: Fair Allocation of Multiple Resource Types}
Dominant Resource Fairness (DRF) generalizes max-min fairness to multi-resource settings by equalizing users’ shares of their dominant resource~\cite{ghodsi2011dominant}.

DRF ensures: \textbf{sharing incentive}, \textbf{strategy-proofness}, \textbf{envy-freeness}, and \textbf{Pareto efficiency}.

\textbf{Example:} With 9 CPUs and 18\,GB RAM:
\begin{itemize}
  \item User A: (1 CPU, 4\,GB) per task, dominant = RAM
  \item User B: (3 CPUs, 1\,GB) per task, dominant = CPU
\end{itemize}
Let $x$, $y$ be the number of tasks for A and B. Constraints:
\[
x + 3y \leq 9,\quad 4x + y \leq 18,\quad \frac{2x}{9} = \frac{y}{3}
\]
Solving yields $x=3$, $y=2$; A uses 3 CPUs and 12\,GB, B uses 6 CPUs and 2\,GB.

\textbf{Scheduling:} DRF selects the user with the lowest dominant share whose task fits. Example sequence:
\begin{itemize}
  \item Step 1: B (dominant share $1/3$)
  \item Step 2: A twice ($2/9 \rightarrow 4/9$)
  \item Step 3: B again until resources are exhausted
\end{itemize}

\textbf{Ant Colony Algorithm}

In ACO-based micro-service scheduling, worker ants simulate task (micro-service) assignment to servers. Each ant starts from a random task, selects servers probabilistically based on pheromone trails, and updates pheromones through frequent traversal (reinforcement) or evaporation (decay). Ants avoid redundant paths and iteratively assign all tasks, collectively optimizing task distribution.

The heuristic integrates three objectives: (1) minimizing inter-service network overhead, (2) balancing cluster load, and (3) reducing average request failure rate. Solution quality is fed back into pheromone updates, enhancing future path selection. This dynamic, intelligent scheduling approach improves reliability, reduces network overhead, balances loads, and lowers carbon emissions~\cite{8744199}.

\subsubsection{AI-Driven and Deep Reinforcement Learning Schedulers}
\textbf{Smart-Kube: Energy-Aware and Fair Scheduler}
The core idea of this paper is to develop \textbf{Smart-Kube}, a Kubernetes-compatible scheduler that uses \textbf{Deep Reinforcement Learning (DRL)} to optimize energy consumption while maintaining fairness in multi-node cluster environments~\cite{ghafouri2023smart}. Traditional schedulers may not effectively balance energy efficiency, fairness, and long-term performance. \textbf{Smart-Kube} addresses this issue by introducing a \textbf{multi-objective reward function} that simultaneously considers energy savings, fairness in task allocation, and long-term system performance.

Smart-Kube's DRL agent learns to allocate tasks across nodes to achieve the following objectives:

\begin{itemize}
    \item Minimize the cluster's energy consumption.
    \item Prevent individual nodes from becoming overloaded to ensure fairness.
    \item Ensure long-term system performance stability by avoiding local optima.
\end{itemize}

\textbf{Smart-Kube} is designed to run within the Kubernetes ecosystem, introducing a DRL-based intelligent decision-maker by replacing or enhancing the default scheduler. This makes it a practical solution for real-world cloud-native deployments aimed at reducing operational energy consumption. It demonstrates how modern AI technology can be seamlessly integrated into Kubernetes to make scheduling smarter, more environmentally friendly, and fairer, aligning with broader goals of sustainable and responsible computing.

\textbf{Deep Learning-Driven Scheduler}
The DL cluster scheduler automatically matches the learning process of the scheduling policy and adapts to changes in the workload and the implementation of the ML framework~\cite{9328612}.

The Deep Learning Scheduler uses a combination of offline supervised and online reinforcement learning. Deep learning improves performance by 44.1\% compared to a fairness scheduler (i.e. DRF)~\cite{peng2021dl2} 

Offline supervised learning: A neural network is trained based on past job data (e.g. carbon emission, job characteristics, resource allocation, and completion time). The network learns to recognize the relationship between these factors so that it can predict how different resource allocations will affect the training time for new jobs.

Online Reinforcement Learning: As new jobs are submitted, the trained network is further adapted based on its performance on these new jobs, allowing the network to continually adapt and improve its prediction of future job scheduling decisions. 

\textbf{MAIZX: A Carbon-Aware Framework for Optimizing Cloud Computing Emissions}
The \textbf{MAIZX} framework provides a scalable, carbon-aware solution that optimizes cloud computing operations by dynamically assessing the real-time and predicted carbon intensity, power usage effectiveness (PUE), and energy efficiency of computing nodes, including private clouds, hybrid clouds, and multi-cloud environments. By directly integrating with virtualization hypervisors and applying an agent-based decision-making mechanism, \textbf{MAIZX} can effectively reduce carbon dioxide emissions by over 85\%, while adapting to fluctuating workloads and carbon emission conditions, ensuring that performance and reliability remain unaffected~\cite{MAIZX}.

\textbf{Greenflow: A carbon-efficient scheduler for deep learning workloads}
\textbf{GreenFlow} addresses the significant carbon emissions generated during deep learning training on GPU clusters by introducing a carbon-aware scheduler. This scheduler dynamically adjusts GPU allocation within a specified carbon budget to minimize the average job completion time (JCT). The system uses performance models to predict task throughput and energy consumption under different configurations and dynamically adjusts GPU allocation based on grid carbon intensity. \textbf{GreenFlow} also employs techniques such as network bundling and peer allocation to optimize resource utilization and reduce emissions caused by fragmentation. Evaluation results demonstrate that \textbf{GreenFlow} can significantly improve performance while maintaining compliance with carbon constraints~\cite{gu2024greenflow}.

\subsubsection{Serverless, Edge, and Federated Solutions}
\textbf{GreenCourier: Carbon-Aware Scheduling for Serverless Functions}
GreenCourier is a scheduling framework that minimizes the carbon footprint of serverless functions by leveraging real-time emissions data~\cite{chadha2023greencourier}.

\textbf{Carbon Data Integration:} It uses sources like WattTime and the Carbon-Aware SDK to monitor regional carbon efficiency in real-time.

\textbf{Geographical Scheduling:} Functions are dispatched to regions with the lowest current emissions, optimizing for carbon efficiency across distributed locations.

\textbf{Carbon-Aware Algorithm:} The scheduler prioritizes deployments that minimize emissions per invocation, reducing the overall carbon impact of serverless computing.

\textbf{CAFE: Carbon-Aware Federated Learning}
Federated Learning (FL) enables distributed training across data centers while preserving data locality, a necessity due to legal and regional constraints. CAFE reduces emissions by selecting data centers in low-carbon regions during each training round~\cite{cafe}.

\begin{itemize}
    \item \textbf{Data Quality Variation:} CAFE introduces a probing step to estimate model gradients using sampled data from candidate data centers.
    
    \item \textbf{Carbon Intensity Uncertainty:} To handle unknown future carbon levels, it uses a Lyapunov Drift-plus-Penalty framework that balances model performance and emission bounds.

    \item \textbf{Time-Constrained Selection:} A greedy algorithm selects an optimal subset of data centers based on convergence speed, test accuracy, utility, and carbon footprint.

    \item \textbf{Double Greedy Algorithms:} Both deterministic and randomized versions iteratively grow and shrink sets of data centers, balancing gain and loss in training performance.
\end{itemize}

\textbf{CEFL: Carbon-Efficient Federated Learning}
CEFL introduces a carbon-aware client selection framework that optimizes cost-to-accuracy by considering each client's carbon emissions, calculated as energy consumed $\times$ average carbon intensity~\cite{cefl}.

\textbf{Prior Strategies:}
\begin{itemize}
    \item \textbf{Random Selection:} Inefficient in accuracy and resource usage.
    \item \textbf{Data-Utility Based:} Clients selected based on contribution to accuracy, with more participants in early training rounds (critical learning periods).
\end{itemize}

\textbf{CEFL Strategy:}
\begin{itemize}
    \item \textbf{Client Cost Awareness:} Accounts for variability in energy and carbon cost (e.g., smartphones vs. data centers).
    \item \textbf{Score Aggregator:} Combines client cost and statistical utility to compute utility-per-cost and selects top clients accordingly.
    \item \textbf{Critical Learning Period:} Uses more clients early to accelerate convergence, then fewer to conserve cost.
\end{itemize}

CEFL reduces carbon emissions by up to 80\%, with only a 38\% increase in training time compared to accuracy-only strategies.

\textbf{KEDA: Kubernetes-based Event Driven Autoscaling}
KEDA is an open-source Kubernetes operator enabling event-driven autoscaling for workloads using sources like Kafka, Prometheus, or AWS SQS~\cite{KEDA}. It integrates with Kubernetes' Horizontal Pod Autoscaler (HPA) by exposing custom metrics and can scale workloads from zero to optimize resource use.

\textbf{How It Works:}
\begin{itemize}
    \item \textbf{Deployment:} Runs as an operator, watching ScaledObjects that define scaling rules.
    \item \textbf{Event Monitoring:} Triggers scaling by evaluating event source metrics.
    \item \textbf{HPA Integration:} Feeds event metrics into HPA for scaling decisions.
    \item \textbf{Scale-to-Zero:} Frees resources by scaling to zero when idle.
\end{itemize}

\textbf{Carbon-Aware KEDA Operator:}
An extension that uses real-time carbon intensity data (e.g., from WattTime or Electricity Map) to limit KEDA scaling. It introduces a custom resource \texttt{CarbonAwareKedaScaler} to set \texttt{MaxReplicaCount}, throttling workloads during high-carbon periods and increasing scale when carbon intensity is low.

\textbf{CASA: A Framework for SLO- and Carbon-Aware Auto scaling and Scheduling in Serverless Cloud Computing}
This paper proposes a carbon emissions and Service Level Objectives(SLOs) aware container scheduling and auto-scaling framework called \textbf{CASA}, designed for serverless computing platforms. The framework aims to address the challenge of balancing carbon footprint minimization with performance maintenance in a Function-as-a-Service (FaaS) environment. Traditional energy-saving methods, such as shutting down idle containers, often result in increased latency due to cold starts, while methods that improve performance by preheating containers increase emissions. \textbf{CASA} achieves this balance by dynamically adjusting container scheduling and auto-scaling strategies based on carbon intensity and SLO constraints. Experimental evaluations show that \textbf{CASA} significantly reduces carbon emissions and service level violation rates compared to existing best practices~\cite{qi2024casa}.

\textbf{GreenWhisk: Emission-Aware Computing for Serverless Platform}
\textbf{GreenWhisk} is a carbon-aware serverless computing platform built on Apache OpenWhisk, designed to reduce the environmental impact of jobs execution in cloud environments. It addresses fluctuations in grid carbon intensity and renewable energy availability by integrating carbon-aware load balancing algorithms. \textbf{GreenWhisk} supports both grid-connected and grid-isolated modes, enabling transparent jobs scheduling based on real-time carbon and energy data without compromising performance. As a flexible infrastructure, the platform can seamlessly integrate various carbon-awareness strategies into serverless architectures~\cite{serenari2024greenwhisk}.

\subsubsection{Carbon Emission Index Priority Scheduling Frameworks}
\textbf{Carbon Emission-Aware Job Scheduling for Kubernetes Deployments}

The goal of this paper is to present a practical approach to reducing carbon emissions in Kubernetes by smartly scheduling workloads. The core idea is to use history energy statistics to identify times when energy has less carbon footprints and shift non-critical tasks to those periods. We make sure this doesn't affect the performance of critical jobs, so services stay reliable~\cite{piontek2024carbon}.

The key contributions include:

\begin{itemize}
    \item A \textbf{carbon emission-aware scheduling algorithm} that integrates seamlessly with Kubernetes by adapting job submission times based on historical carbon intensity trends.
    \item A \textbf{priority-aware mechanism} to distinguish between critical and delay-tolerant jobs, ensuring that only flexible workloads are shifted to reduce emissions.
    \item Demonstration of substantial \textbf{carbon footprint reduction} with negligible impact on job performance in practical Kubernetes clusters.
\end{itemize}

This work shows how Kubernetes can be adapted to make cloud computing greener. By taking into account periods of lower energy consumption, we can run workloads more sustainably without sacrificing speed or system availability.

\textbf{CASPER: Carbon-Aware Scheduling and Provisioning for Distributed Web Services}
CASPER reduces the carbon footprint of geographically distributed web services by scheduling requests in regions with lower carbon intensity, while meeting service-level latency constraints~\cite{Souza_2023}. Deployed in Kubernetes, CASPER consists of:

\begin{itemize}
    \item \textbf{Carbon-Aware Provisioner (CAP):} Optimizes the number and location of active servers to minimize emissions while satisfying latency SLOs. It formulates a multi-objective problem using estimated carbon intensities and user request data, ignoring resource and communication constraints except server count.
    
    \item \textbf{Carbon-Aware Scheduler (CAS):} Acts as a load balancer, distributing user requests across regions per CAP's provisioning plan, and schedules requests not covered by CAP.
\end{itemize}

Experiments show up to 70\% energy savings with minimal performance impact when CASPER is integrated with Kubernetes.

\textbf{PCAPS: Carbon- and Precedence-Aware Scheduling for Data Processing Clusters}
This paper explores how to address carbon-aware scheduling challenges in large-scale data processing tasks where task dependencies exist. It introduces \textbf{PCAPS}, a scheduler that combines time-varying carbon intensity data with task priorities to avoid delaying critical upstream tasks that could block the entire processing pipeline. \textbf{PCAPS} integrates with machine learning based schedulers and provides a configurable trade-off between minimizing carbon emissions and maintaining task completion times. The authors also propose CAP, a generic wrapper for carbon-neutral schedulers that inherits \textbf{PCAPS}'s resource scheduling strategies. Experiments on a 100-node Kubernetes Spark cluster demonstrate that \textbf{PCAPS} can reduce carbon emissions by up to 32.9\% while maintaining overall scheduling efficiency~\cite{lechowicz2025pcaps}.

\textbf{Operating Cloud Applications Under a Carbon Budget}
This paper proposes a carbon-aware deployment strategy for long-running microservice-based cloud applications that maintains application availability and performance under all conditions. Unlike traditional carbon-aware scheduling methods for batch tasks, this method operates on an hourly carbon budget and dynamically adjusts application configurations (including microservice versions and horizontal scaling) to optimize user quality of experience(QoE) and revenue while meeting environmental constraints. The authors propose an optimization algorithm that selects the most appropriate deployment configuration based on workload, carbon intensity, and budget, and validate its effectiveness through a flight booking application. The proposed method outperforms the baseline in terms of QoE and revenue while meeting the carbon budget, demonstrating its practical application value in real-world dynamic environments~\cite{espinosa2024optimizing}.

\textbf{U-DUCT: Uncertainty-aware Dynamic Unified Carbon Modeling Tool for Datacenter Scheduling}
The \textbf{U-DUCT} framework proposes an uncertain, dynamic, and unified carbon emissions modeling tool designed to comprehensively assess carbon emissions in data centers during both the design phase and operational phase. Unlike previous tools, \textbf{U-DUCT} not only considers computing servers but also incorporates the contributions of often-overlooked components such as storage devices and switches. Additionally, the framework accounts for variability and uncertainty in hardware and software characteristics, which can impact embedded carbon emissions and operational carbon emissions. Through this holistic and runtime-aware model, \textbf{U-DUCT} has identified new opportunities for emissions reduction in the actual computational workloads of data center components~\cite{guan2024u}.

\textbf{A Green Cloud-Based Framework for Energy Efficient Task Scheduling Using Carbon Intensity Data for Heterogeneous Cloud Servers}

This paper proposes a cloud-based scheduling framework that integrates real-time carbon intensity data to optimize the execution of energy intensive tasks in cloud data centers. The framework applied Kubernetes, AWS services, and containerized workloads to dynamically reschedule jobs with high energy consumptions based on the changes of regional carbon intensity. It results reducing operational carbon emissions without compromising performance. The framework emphasizes automation, scalability, and compatibility with major cloud providers, supporting sustainable practices aligned with the United Nations' Sustainable Development Goals. The system enhances energy efficiency through intelligent workload management to address the growing energy demands of artificial intelligence and machine learning applications~\cite{beena2025green}.

\textbf{LinTS: Carbon-Aware Temporal Data Transfer Scheduling Across Cloud Data centers}
\textbf{LinTS} is a carbon-aware scheduling framework designed to reduce the environmental impact of inter-data center communication, which is the primary source of carbon emissions in cloud computing. By leveraging the temporal variation in carbon intensity across different data centers, \textbf{LinTS} optimizes data transmission scheduling to ensure that transfers occur during low-carbon periods while meeting all transmission deadlines. The system outperforms traditional heuristic algorithms through intelligent scaling and transmission decisions, reducing carbon emissions by up to 66\% compared to worst-case scenarios and by 15\% compared to state-of-the-art baselines. \textbf{LinTS} demonstrates how to effectively utilize time-based carbon data to enhance the sustainability of cloud infrastructure without compromising operational constraints~\cite{rodrigues2025carbon}.

\section{Challenges and Research Opportunities} \label{sec:future}

\subsection{GPU Scheduler}
Enhancing Kubernetes scheduling with AI, particularly for GPU resource management, is a promising area. Song proposes dividing physical GPUs into multiple virtual GPUs to boost cluster utilization by 10\%~\cite{song2018gaia}. Thinakaran develops a Kubernetes scheduler leveraging real-time GPU usage metrics to dynamically allocate GPU resources~\cite{thinakaran2019kube}. However, GPU scheduling in Kubernetes remains underexplored, with efficient container-level resource management still needing advancement~\cite{carrion2022kubernetes}.

\textbf{SCI Data and Location-Shifting} Improving software carbon intensity (SCI) data is essential. Collecting energy data for virtual machines is harder than for bare metal. Additionally, configuring carbon intensity based on both time and location could offer more effective scheduling opportunities.

\subsection{Algorithm Trade-offs}
Current schedulers prioritize resource utilization but often ignore power usage efficiency (PUE), a key metric for sustainability. Reducing carbon intensity may require trade-offs, such as choosing between low-latency, high-carbon regions and high-latency, low-carbon ones. Future algorithms must account for the dynamic nature of both workloads and energy profiles.

\subsection{Microservices Scheduling}
Kubernetes faces challenges in managing microservices and modeled workloads. Pod-level scheduling can be inefficient for microservices. Gang scheduling, which groups related pods, can improve coordination. Meanwhile, workload prediction remains underdeveloped and is vital for efficient scheduling under dynamic demands.

\subsection{GPU Power Capping for Sustainable AI}
GPU power capping can improve energy efficiency in HPC. A study on MIT SuperCloud showed that applying a 60\% power cap significantly reduced power consumption and GPU temperatures with minimal performance impact. However, care is needed to avoid triggering excess jobs that offset energy savings. Integrating such capping techniques into Kubernetes could benefit sustainable AI workloads~\cite{zhao2023sustainable}.

\subsection{Data Center Optimization}
Kubernetes can reduce carbon intensity through component impact measurement, task rescheduling, and carbon-aware algorithms. Baris shows that eco-friendly UPS batteries lower peak power costs~\cite{baris}, while Kontorinis finds no performance loss from such substitutions~\cite{kontorinis2012managing}. Kuroda enhances power systems by reducing AC-to-DC conversions~\cite{kuroda2013high}. Hancock recommends cold-region data centers (e.g., Iceland) for cooling benefits~\cite{HancockIceland}. Future modeling should examine how power delivery and consumption influence carbon intensity~\cite{datacentersreview}.

\subsection{Federated Learning}
One of the main challenges facing FL algorithms is the heterogeneity of client data. For example, different types of data may not be independently and identically distributed (non-IID) in terms of distribution, volume, or quality. For instance, some devices may only see digits 0 and 1 with skewed frequencies~\cite{nonIIDexplain}, affecting aggregation and model convergence.

This problem worsens when FL is deployed on edge devices using green energy (e.g., wind turbines or sensors), which have variable data volumes due to memory, load, and location differences~\cite{brecko2022federated,xie2022improving}. Limited bandwidth and intermittent connectivity further disrupt timely model updates~\cite{kang2020reliable}. Designing robust aggregation strategies under these constraints is an emerging research trend~\cite{abreha2022federated}.

\section{Conclusion} \label{sec:conclusion}

This report discusses the shift in cloud computing from virtualization to containerization and the energy consumption challenges associated with these technologies, with a focus on the industry-standard container orchestration engine, Kubernetes. This work investigates not only the default scheduling and auto-scaling features of Kubernetes but also advanced hardware and software optimizations that can improve energy efficiency and carbon awareness in large scale data center operations. In our extensive literature review, we discuss the optimizations that have been applied to the Kubernetes scheduler for temporal and spatial shifting of workloads which reduces the environmental impact of workload execution. This report investigates the use of state-of-the-art machine learning techniques such as deep reinforcement learning and federated learning for optimized workload scheduling in Kubernetes. We emphasize enhancing carbon efficiency in federated learning workloads within data centers, while also leveraging federated learning to optimize the scheduling of distributed workloads without breaking data privacy by training models in a decentralized manner. The analysis of the challenges and research opportunities highlights that future research in this direction necessitates close cooperation between the government and enterprises, universities and other organizations. The development of reasonable standards and norms will also help promote the promotion and application of this technology.

\bibliographystyle{IEEEtran}
\bibliography{thebibliography}

\end{document}